\documentclass[11pt]{article}

\usepackage{jheppub}

\usepackage{amsmath,amsfonts,amsbsy,amssymb,array,accents,dsfont}
\usepackage{enumerate,array,latexsym,graphicx,mathrsfs,verbatim,psfrag}
\usepackage{bm} 
\usepackage[normalem]{ulem}
\usepackage{booktabs} 
\usepackage{chngpage} 

\usepackage{enumitem}
\setlist{itemsep=0pt}

\newcommand{\captionfonts}{\small}
\makeatletter  
\long\def\@makecaption#1#2{%
  \vskip\abovecaptionskip
  \sbox\@tempboxa{{\captionfonts #1: #2}}%
 \ifdim \wd\@tempboxa >\hsize
    {\captionfonts #1: #2\par}
  \else
    \hbox to\hsize{\hfil\box\@tempboxa\hfil}%
  \fi
  \vskip\belowcaptionskip}
\makeatother   

\DeclareMathSymbol{\medhatsym}{\mathord}{largesymbols}{"62} 

\DeclareMathSymbol{\medtildesym}{\mathord}{largesymbols}{"65}

\numberwithin{equation}{section} 

\makeatletter
\g@addto@macro\bfseries{\boldmath}
\makeatother

\def\coeff#1#2{\relax{\textstyle {#1 \over #2}}\displaystyle}

\def\IR{\mathds{R}}

\def\cA{{\cal A}}
\def\cB{{\cal B}}
\def\cC{{\cal C}}

\def\cF{{\cal F}}

\def\cK{{\cal K}}

\def\cM{{\cal M}}
\def\cN{{\cal N}}
\def\cO{{\cal O}}
\def\cP{{\cal P}}
\def\cQ{{\cal Q}}
\def\cR{{\cal R}}
\def\cS{{\cal S}}

\def\cO{{\cal O}}

\catcode`,\active

\catcode`\,12

\newcommand{\eqn}[1]{\begin{align}\begin{split}#1\end{split}\end{align}}

\def\tight#1{\! #1 \!}

\def\({\left(}
\def\){\right)}
\def\[{\left[}
\def\]{\right]}

\def\ie{{ \it i.e.}}

\def\th{{\rm th}}

\def\tauhat{\tau}
\def\tauf{\tauhat_{\! f}^{~}}

\def\half{\frac12}
\def\coeff#1#2{{\textstyle \frac{#1}{#2}}}
\def\hf{\coeff12}

\def\One{{\hbox{1\kern-1mm l}}}

\def\barray{\begin{array}}
\def\earray{\end{array}}
\def\be{\begin{equation}}
\def\ee{\end{equation}}
\def\bea{\begin{eqnarray}}
\def\eea{\end{eqnarray}}
\def\bal{\begin{align}}
\def\eal{\end{align}}
\def\nn{\nonumber}

\def\IR{\mathbb{R}}
\def\IS{\mathbb{S}}
\def\IT{\mathbb{T}}

\begin{document}
\begin{flushright}
~\\
\end{flushright}

\vspace{-.5cm}

\begin{center}

{\LARGE {\bf 
The Harder They Fall, the Bigger They Become: \\[.1cm]
Tidal Trapping of Strings by Microstate Geometries \\[.3cm]
}}

\vspace{9mm}

{\large
\textsc{Emil J. Martinec$^{1}$ and Nicholas P.~Warner$^{2,3}$}}

\vspace{8mm}

$^1$Enrico Fermi Inst.\ and Dept.\ of Physics, \\
University of Chicago,  5640 S. Ellis Ave.,
Chicago, IL 60637-1433, USA\\
\medskip
$^2$Institut de Physique Th\'eorique,
Universit\'e Paris Saclay,\\
CEA, CNRS, F-91191 Gif sur Yvette, France \\
\medskip
$^3$Department of Physics and Astronomy
and Department of Mathematics,\\
University of Southern California,
Los Angeles, CA 90089, USA

\vspace{4mm} 
{\footnotesize\upshape\ttfamily  ejmartin @ uchicago.edu, warner @ usc.edu} \\

\vspace{11mm}
 
\textsc{Abstract}

\end{center}

\begin{adjustwidth}{12mm}{12mm} 
 
\vspace{-2mm}
\noindent
We consider the fate of a massless (or ultra-relativistic massive) string probe propagating down the BTZ-like throat of a microstate geometry in the D1-D5 system.  Far down the throat, the probe encounters large tidal forces that stretch and excite the string.  The  excitations are limited by the very short transit time through the region of large tidal force, leading to a controlled approximation to tidal stretching. We show that the amount of stretching is proportional to the incident energy, and that it robs the probe of the kinetic energy it would need to travel back up the throat.  As a consequence, the probe is effectively trapped far down the throat and, through repeated return passes,  scrambles into the ensemble of nearby microstates.  We propose that this tidal trapping may lead to weak gravitational echoes.

\end{adjustwidth}

\thispagestyle{empty}
\newpage


\baselineskip=14pt
\parskip=2pt

\tableofcontents


\baselineskip=15pt
\parskip=3pt

\vskip 2cm
\hrule
\vskip 1cm

\section{Introduction and Summary}
\label{Sect:introduction}

Black holes are the ultimate particle accelerators -- according to classical effective field theory, infalling probes encounter arbitrarily violent forces at the black hole singularity, where the spacetime geometry creates tidal forces that become arbitrarily large.  In string theory, probes of the background are extended objects, which are expected to be tidally disrupted as they approach the singularity.  One might expect this disruption to signal the beginning of the probe's scrambling into the space of black-hole microstates.  Indeed, there is considerable evidence that the black hole itself is a complicated bound state of such extended objects (see for instance~\cite{Maldacena:1996ky,Horowitz:1996rn,Martinec:1999bf,Peet:2000hn,Mathur:2005zp,Bena:2007kg,Bena:2013dka}
for reviews, and also~\cite{Maldacena:1996ya,Horowitz:1996ay,Danielsson:2001xe}).

Quantum effects lead to puzzles such as the black-hole information paradox. There is a substantial body of evidence that resolving this paradox requires new physics at the horizon scale.  Perhaps the most conservative (and to some extent computationally accessible) proposal is that string theory gives rise to this new horizon-scale structure.  For instance,  qualitative estimates in the BFSS matrix model  \cite{Banks:1997hz,Banks:1997tn,Horowitz:1997fr} indicate that the wavefunction of the black hole constituents extends out to the horizon scale, and the string/black-hole correspondence transition~\cite{Horowitz:1996nw,Martinec:1999bf,Damour:1999aw} indicates that a conventional brane bound state emerges from the horizon as the coupling is dialed down.  It has also been shown that, as the string coupling increases, some brane bound states grow at the same rate as the horizon scale and that such structures can support microstructure and prevent the formation of a horizon (see, for example, \cite{Bena:2004wt, Bena:2004wv,Bena:2007kg,Bena:2013dka}).

If such a stringy resolution is built entirely from structures available within supergravity, then it must be some form of microstate geometry \cite{Gibbons:2013tqa}.   Put differently, microstate geometries provide the only possible, fully back-reacted mechanism for supporting a rich variety of microstate structure at the horizon scale, within the supergravity approximation.  Regardless of whether or not such microstate geometries play a significant role in the complete phase space of black-hole microstates, the important point is that such geometries provide an invaluable laboratory for studying and probing horizon-scale microstructure.  This will be their role in this paper.

Probes of microstate geometries have  been extensively studied \cite{Tyukov:2017uig,Raju:2018xue,Bianchi:2018kzy,Bena:2018mpb,Bena:2019azk,Bena:2020iyw} and this has revealed several black-hole like features.  In particular, the geometric structure closely resembles that of black hole until one  is extremely close to the horizon scale. This means that, on short time-scales, the Green functions  exhibit thermal decay \cite{Bena:2019azk}. However, these geometries also differ significantly from black-holes, and, most significantly, the fact that geometry is capped off at a large, but finite, redshift means that information about the state of the solution can be probed from infinity.

One of the surprises that emerged from these probe calculations was that the small deviations from the black-hole geometry could give rise to extremely large (Planck/string/compactification scale) tidal forces within the throat of the geometry, long before the cap is reached \cite{Tyukov:2017uig,Bena:2018mpb,Bena:2020iyw}.  This suggested that probes would actually be tidally disrupted before they encounter the cap of the microstate geometry.  Our goal here is to analyze this phenomenon using string probes.  We encounter another very interesting surprise: While the string does indeed encounter tidal forces that are strong enough to overcome the string tension, these tidal forces act over a very short proper time and result in string mode excitations of $\cO(1)$, up to a mode cutoff set by the incident energy.
Thus, if the string enters as a massless particle, it emerges in a massive state, traveling on a time-like geodesic.  As with any tidal excitation, the energy for the excitation comes from the kinetic energy of the probe, and so the tidal interaction causes the particle to become more deeply trapped in the geometry.

\begin{figure}[h!]
\centering
\includegraphics[width=.58\textwidth]{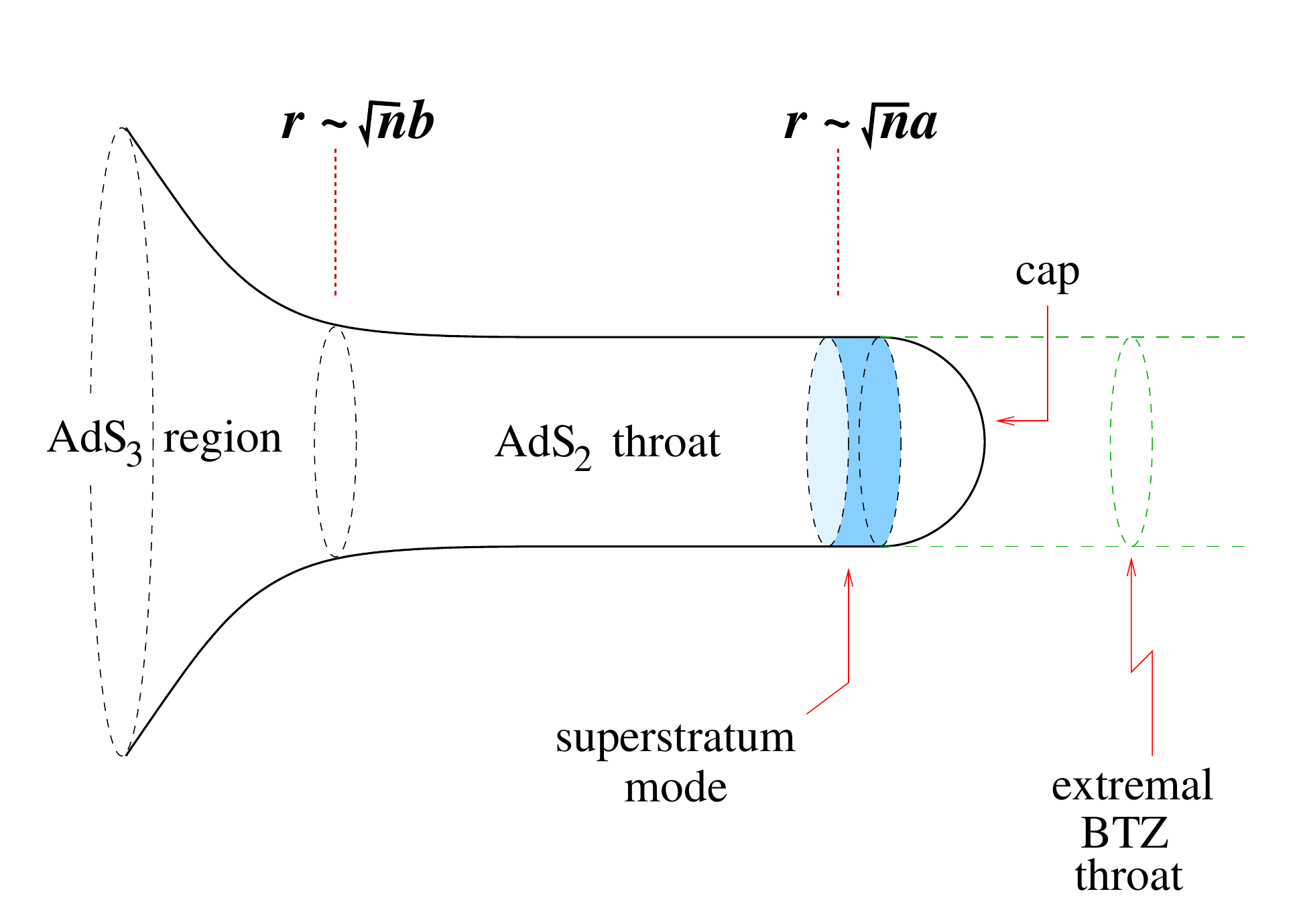}
\caption{\it The structure of the superstratum microstate geometry that we probe with an infalling string, and its relation to the rotating BTZ geometry.}
\label{fig:Superstratum}
\end{figure}

In order to elucidate this behavior in a little more detail, we note that there are two scales,  $a$ and $b$,  and a  positive integer parameter, $n$ that characterize the background.   The scales $a$ and $b$ are, respectively,  set by the $\IS^3$ angular momentum and the D1 and D5 charges of the background.  The parameter $n$ sets the scale of the momentum charge relative to the geometric mean of the D1 and D5 charges.   The scale $r\sim \sqrt{n}\,b$ characterizes the top of the BTZ-like throat, where the asymptotically $AdS_3$ geometry transitions to a circle fibration over $AdS_2$. The ratio  $b/a$ determines the redshift down the BTZ throat, and this is assumed to be extremely large.  The geometric cap corresponds roughly to the region $r \lesssim \sqrt{n} \,a$;   The microstate geometry, and the way it approximates the BTZ geometry, are depicted in Figure~\ref{fig:Superstratum}. 
\vspace{0.5mm}
\vspace{1mm}

It was shown in \cite{Tyukov:2017uig,Bena:2018mpb} that for a particle falling from $r \sim b$, the tidal forces become ``large''  (Planck/string/compactification scale) at about $r \sim \sqrt{ab}$.  These forces arise from the non-trivial multipoles of the background, magnified by the relativistic speed of the infalling probe.   However, as we will show, the proper time over which the string probe encounters large tidal forces is extremely short, and limits the effect.  In the end, a finite fraction of the string's initial center-of-mass energy is converted to string oscillator excitations as it passes through the cap.

We will show that a key effect of the tidal forces near the cap is to trap the string, and that each pass through the cap excites the string further, until all its center-of-mass energy is converted into stringy excitations.  We will calculate the effect of tidal forces on the first pass down the throat, and find that after reflecting off the cap the string returns to a maximum radial position of order
\be
\label{rmax intro}
r_{\rm max}   ~\sim~   \Big( a^3 b\, n^3 g_6^{\,2}\, \sqrt{j_L} \, \Big)^{\frac14}   
\ee
where  $j_L$ is a quantized $\IS^3$ angular momentum, carried by the microstate geometry; $g_6$ is the string coupling scaled by the compactification radius.  The string thus never returns to the height from which it came, and in fact it never rises to the top of the throat at $r\sim b$, or even the location $r\sim \sqrt{ab}$ where the tidal forces first get large, even if it was sent in from much further out. Instead, it must ultimately settle  into an excited string state in the bottom of the cap of  the microstate geometry.  Interestingly, $r_{\rm max}$ is independent of the incident energy $E$ of the string; increasing $E$ simply increases the number of string oscillator modes that get excited, out to a maximum mode number
\be
k_{\rm max} \sim \frac{b\,\alpha'  E}{a^2n} ~,
\ee
with each tidally excited mode having an excitation level of order one.  This means that a fixed fraction of the overall energy is transferred from center of mass motion to string excitations and, as a result,  the return height  (\ref{rmax intro}) is independent of $E$.

This behavior has several important consequences for the microstate geometry program.  Probe calculations, such as that in~\cite{Bena:2019azk}, exhibit a ``sharp echo'' in a return pulse at time scales of order $b^2/a^2$ times the $AdS$ scale $R_{AdS}$.  This return time-scale can be as large as the central charge of the CFT, and while that is rather long, it is nowhere near long enough for black-hole behavior.  The time delay, and the sharpness of return echoes, appeared to be an artifact of a rigid supersymmetric geometry in a highly specialized, coherent, zero-temperature state.  It was believed that in order to see more chaotic, long-term trapping behavior (such as that suggested in  \cite{Eperon:2016cdd, Marolf:2016nwu}), one would have to incorporate non-trivial back-reaction in which the probe loses energy to the background, or through the emission of radiation.   Our results show that the trapping of probes can arise from a simpler, intrinsically stringy mechanism.  While analysis using geodesics and wave equations produces returns and echoes, these should be viewed as approximations to string probes, and we show that a string probe will never return.  Indeed, even for  zero-temperature (supersymmetric), highly coherent microstate geometries, string probes beome trapped in the cap, so long as the throat geometry is deep enough to generate large tidal forces.   

This is not to say that radiation by probes, or energy transfer to the background are unimportant.  Indeed these will be an essential parts of the scrambling process.  
And as we will discuss in Section \ref{sec:Conclusions}, it is still possible that the tidal trapping of a probe could still result in some form of low-energy echo. 
The beauty of the result presented here is that it exhibits an important trapping mechanism that can be analyzed in a controlled approximation without back-reaction.

We now proceed to details of the computation of string probe dynamics in microstate geometries.  
In Section~\ref{sec:MG}, we review the structure of microstate geometries, and specialize to the vicinity of a radially ingoing null geodesic followed by a massless (or highly relativistic) probe.  To simplify the analysis of probe string dynamics, we take the Penrose limit.  String propagation in this limit is solvable in a light-cone gauge~\cite{Horowitz:1990sr}, where the worldsheet dynamics reduces to a free field problem with a time-dependent mass matrix.  We analyze this problem in Section~\ref{sec:stringprop} in the WKB approximation, and estimate the Bogoliubov coefficients for the string's oscillator modes as it reaches the cap of the geometry.  Having estimated the mass of the resulting string, we return to the full geometry to estimate the properties of the now timelike trajectory of the string, and in particular the maximum height~\eqref{rmax intro} it reaches as it tries to return up the throat.  In Section~\ref{sec:Tides}, we re-examine geodesic deviation in the throat in order to provide a heuristic explanation of these results.  In Section \ref{sec:Conclusions} we discuss the broader impact of our results for the scrambling and relaxation of probes into the microstructure of black holes.
 
\section{The Microstate Geometry}
\label{sec:MG}

We are going to focus on one particular family of microstate geometries, namely the $(1,0,n)$ superstrata.  This family  has the  advantage of  being relatively simple and computable and, as a result, has been widely analyzed (see, for example, \cite{Bena:2017upb,Tyukov:2017uig,Raju:2018xue,Bianchi:2018kzy,Heidmann:2019zws,Bena:2019azk,Heidmann:2019xrd,Mayerson:2020tcl}).  As we will discuss in Section \ref{sec:Conclusions}, we expect our results to be far more universal, applying to all superstrata, and probably to all microstate geometries, and perhaps to any  coherent geometric realization of black-hole microstructure.

\subsection{The geometry}
\label{ss:10nmetric}

The simplest way to describe this geometry is to use the form given in \cite{Bena:2017upb} and simplify it using the more recent insights of \cite{Heidmann:2019xrd,Mayerson:2020tcl}.   That is we write the geometry as a non-trivial, deformed $S^3$ fibration over a three-dimensional base manifold, $\cK$, that is asymptotic to AdS$_3$.  To do this, and for future reference, it is useful to introduce some ``bump'' functions and parameters: 
\begin{equation}
F ~\equiv~  1 - \frac{r^{2n}}{(r^2+a^2)^{n}} \,, \qquad   G ~\equiv~ 1 - \frac{a^2 \, b^2}{2 a^2+ b^2} \, \frac{ r^{2n}}{(r^2+a^2)^{n+1}}    \,, \qquad    A ~\equiv~ \sqrt{ 1 +  \frac{b^2}{2 a^2}} 
\label{FGAdefns}
\end{equation}
\begin{equation}
\Lambda ~\equiv~ \sqrt{ 1 - \frac{a^2\,b^2}{(2 a^2 +b^2)} \, \frac{r^{2n}}{(r^2 +a^2)^{n+1}} \, \sin^2 \theta  }   \,, \qquad  \Gamma ~\equiv~ \sqrt{ 1 - \frac{b^2}{(2 a^2 +b^2)} \, \frac{r^{2n}}{(r^2 +a^2)^{n}}   }  
\label{GamLamdefns}
\end{equation}
The six-dimensional metric may then be written
\begin{equation}
ds_6^2 ~=~ \sqrt{Q_1 Q_5} \, \Big(\, \widehat{ds}_3^2 ~+~ \widetilde{ds}_3^2 \,\Big)
\label{sixmet1}
\end{equation}
where
\begin{equation}
 \widehat{ds}_3^2  ~=~ \Lambda \, \bigg[ \frac{dr^2}{r^2 + a^2} ~+~\frac{2\,r^2(r^2 + a^2)}{R_y^2 \, a^4}  \, dv^2 ~-~\frac{1}{2\,  R_y^2} \,\frac{1}{A^4  G^2} \, \bigg( du + dv +  \frac{2   A^2 r^2}{a^2} \, dv\bigg)^2\bigg]  \,, 
\label{Kmet1}
\end{equation}
and
\begin{equation}
\begin{aligned}
 \widetilde{ds}_3^2  ~=~ \Lambda \, d\theta^2 & ~+~  \frac{1}{\Lambda}\,  \sin^2 \theta \, \bigg(d \varphi_1  -  \frac{1}{\sqrt{2}\, R_y A^2}\,(du+dv) \bigg)^2 \\
 & ~+~  \frac{G}{\Lambda}\,  \cos^2 \theta\,  \bigg(d \varphi_2  +  \frac{1}{\sqrt{2}\, R_y\,a^2  A^2 \, G}\,\big(a^2(du-dv) - b^2 F \, dv \big) \bigg)^2 \,.
 \end{aligned}
\label{Smet1}
\end{equation}
As usual, $u$ and $v$ are null coordinates, which are related to the canonical time and spatial coordinates via:
\begin{equation}
  u ~=~  \coeff{1}{\sqrt{2}} (t-y)\,, \qquad v ~=~  \coeff{1}{\sqrt{2}}(t+y) \,, \label{tyuv}
\end{equation}
where $y$ is the coordinate around $S^1$ with
\begin{equation}
  y ~\equiv~  y ~+~ 2\pi  R_y \,. \label{yperiod}
\end{equation}

Apart from $R_y$, there are five parameters that define this background: the D1 and D5 brane charges, $Q_1$ and $Q_5$, two real parameters $a$ and $b$ and an integer, $n$, which determine the angular momenta and the momentum charges:
\begin{equation}
\label{eq:ConsCharges}
 J_L  ~=~  J_R ~=~   \frac{R_y}{2} \,a^2\,,\qquad Q_P = \coeff{1}{2 } \, n \, b^2\,.
\end{equation} 
Regularity of the solution requires the constraint:
\begin{equation}
Q_1 Q_5 ~=~  \bigg(a^2 ~+~ \frac{b^2}{2} \bigg) \, R_y^2 \,.
\label{SSreg1}
\end{equation}
The supergravity solution is also supported by fluxes and one can find the precise expressions, and all the relevant details in earlier work, like \cite{Bena:2017xbt,Heidmann:2019zws,Heidmann:2019xrd}.  We will need the details of one of these fluxes and these are given in Appendix \ref{app:Bfield}.

\subsection{The quantized charges}
\label{ss:quant}

Since we are going to consider the detailed dynamics of strings moving in this background, we need to catalog the connection between the supergravity parameters and the quantized charges, the string coupling, $g_s$, the square of the string length, $\alpha'$, and the volume of the four-dimensional manifold, $V_4$.  To do this, one starts by defining
\begin{equation}
\cN ~\equiv~ \frac{V_4\, R_y^2}{ (2\pi)^4 \,g_s^2 \,\alpha'^4}~=~\frac{V_4\, R_y^2}{(2\pi)^4 \, \ell_{10}^8} ~=~\frac{{\rm Vol} (T^4) \, R_y^2}{ \ell_{10}^8} \,.
\label{cNdefn}
\end{equation}
The quantized charges are then given by:
\begin{equation}
\label{eq:quantizedcharges}
 \begin{aligned}
& n_1  ~=~  \frac{V_4}{(2\pi)^4\,g_s\,\alpha'^3}\,Q_1 \,, \quad    n_5  ~=~    \frac{1}{g_s\,\alpha'}\,Q_5 \,,  \\ & n_p ~=~ \coeff{1}{2} \,\cN\, n\, b^2   ~=~    \frac{R_y^2 \,n_1 n_5}{Q_1 Q_5}\, Q_p  
\,, \qquad j_L ~=~ j_R~=~\coeff{1}{2} \, \cN\, a^2\,.
\end{aligned}
\end{equation}
Of particular importance is the following identity between dimensionless ratios of length scales:
\begin{equation}
\frac{a\, R_y}{\alpha'}  ~=~  \bigg(\frac{V_4}{( 2\pi)^4 \alpha'^2 }  \bigg)^{-\frac{1}{2}} \,g_s^2 \,  \sqrt{ j_L }  
~\equiv~ g_6^{\,2}\sqrt{ j_L }   
\label{jLbound}
\end{equation}
where we have defined the six-dimensional string coupling, $g_6$.
In scaling geometries, one takes $a$ to be small, with a lower limit of $ j_L = \frac{1}{2}$.  This identity therefore determines the lower bound on $a R_y$ in terms of $g_6$ and the string length $\sqrt{\alpha'}$.

\subsection{Radial geodesics}
\label{ss:geodesics}
\def\geoE{\hat E}

Since the six-dimensional metric is independent of $(u,v,\varphi_1, \varphi_2)$, this means that the corresponding momenta are conserved:
\begin{equation}
L_1 ~=~ {K_{(1) M }} \frac{dz^M}{d \lambda} \,, \qquad L_2 ~=~ {K_{(2)  M }}  \frac{dz^M}{d \lambda} \,,\qquad  P ~=~ {K_{(3)   M }}  \frac{dz^M}{d \lambda} \,, \qquad E ~=~ {K_{(4)   M }}  \frac{dz^M}{d \lambda}   \,,
  \label{ConsMom}
\end{equation}
where $z^M =(u,v,r,\theta,\varphi_1,\varphi_2)$ and  the $K_{(I)}$  are the Killing vectors: $K_{(1)}  = \frac{\partial}{\partial \varphi_1}$, $K_{(2)}  = \frac{\partial}{\partial \varphi_2}$, $K_{(3)}  = \frac{\partial}{\partial v }$ and $K_{(4)}  = \frac{\partial}{\partial u}$.   There is the standard quadratic conserved quantity coming from the metric: 
\begin{equation}
g_{MN} \, \frac{dz^M}{d \lambda} \, \frac{dz^N}{d \lambda} ~=~ -m^2  \,,
  \label{MetInt}
\end{equation}
where $m =0$ for null geodesics and $m\ne 0$ for time-like geodesics.  It is canonical to take $m = 1$ for time-like geodesics and then $\lambda$ is the proper time measured on the geodesic.   

For future clarity, we note that in general relativity it is standard practice to take the affine parameter, $\lambda$, to have the dimensions of length.   This makes the conserved quantities, $E, P, L_1$ and $L_2$ dimensionless and thus  should be thought of as energy, or angular momenta per unit mass.  In  string theory, the natural choices of parameters on the worldsheet, and hence the affine parameters, are dimensionless, and so the conserved quantities have dimensions  of length.  In particular, $E$, as defined in (\ref{ConsMom}) is a length.  For strings one should replace $E \to \alpha' E$, where $\sqrt{\alpha'}$ is the string length.  This replacement means that $E$ has the correct dimensions and does indeed represent the center of mass energy of the string.  To avoid potential confusion, we will henceforth denote the dimensionless energy of geodesic motion by $\hat E$ and use $\alpha' E$ in the string context.

In \cite{Bena:2017upb} it was shown that there is also a conformal Killing tensor, which leads to the following conserved quantity on null geodesics:
\begin{equation}
\Xi ~\equiv~ \xi_{MN} \, \frac{dz^M}{d \lambda} \, \frac{dz^N}{d \lambda} ~\equiv~Q_1 Q_5 \, \Lambda^2 \, \bigg(\frac{d\theta}{d \lambda}\bigg)^2 ~+~ \frac{L_1^2}{\sin^2 \theta}~+~ \frac{L_2^2}{\cos^2 \theta} \,.
  \label{ConKtens}
\end{equation}
However, here we are going to simplify things by observing that, as a consequence of the discrete symmetries of the metric under $\theta \to - \theta$  and $\theta \to \pi - \theta$, there are geodesics that have $\theta \equiv 0$  or $\theta \equiv \frac{\pi}{2}$  along their entire length.    We will restrict our attention to the former as it makes the analysis significantly easier.   

For radial motion  one must  remove all the centripetal barriers and this means one must take:
\begin{equation}
L_1 ~=~  0\,, \qquad L_2 ~=~  0 \,, \qquad P ~=~ \geoE  \,.
\label{CentBarr1}
\end{equation}
Because of the non-trivial fibration structure  or the metric, the constraint (\ref{CentBarr1})  does not mean that the velocities along $v$, $\varphi_1$ and $\varphi_2$ vanish.  These angular velocities are determined by (\ref{CentBarr1}), and are proportional to $\frac{du}{d \lambda}$.

Since we have taken:
\begin{equation}
\theta ~=~ 0 \,, \qquad \frac{d\theta}{d \lambda}  ~=~ 0 \,,
  \label{thetavel}
\end{equation}
the last remaining velocity, $\frac{d r}{d \lambda}$, is determined by   (\ref{MetInt}), which yields:
\begin{equation}
\bigg(\frac{d r}{d \lambda} \bigg)^2  ~=~ 2\, \geoE^2  A^2 \, \Gamma^2 ~-~ m^2 \, \frac{(r^2 +a^2)}{\sqrt{Q_1 Q_5}}\,.
  \label{radvelsq}
\end{equation}
For  null geodesics, this reduces to 
\begin{equation}
 \frac{d r}{d \lambda}  ~=~ \pm \sqrt{2}\, \geoE A \,\Gamma \,,
  \label{infallnull}
\end{equation}
where $A$ and $\Gamma$ are defined in (\ref{GamLamdefns}).

For our analysis, it is extremely useful to re-parametrize the metric by replacing the coordinate $r$ by the parameter $\lambda$ along a null geodesic.  We fix the normalization of $\lambda$ by setting $\geoE = \frac{1}{2\sqrt{2}}$ and replace $(u,v)$ by $(t,y)$ according to (\ref{tyuv}).   This means that we make the following replacement in the coordinate differentials:
\begin{equation}
\begin{aligned}
 (du,dv,dr,d\theta, d\varphi_1,d\varphi_2) ~\to~  & \Big(\frac{du}{d \lambda},\frac{dv }{d \lambda} ,\frac{dr}{d \lambda} ,\frac{d \theta}{d \lambda} ,\frac{d\varphi_1}{d \lambda} ,\frac{d\varphi_2}{d \lambda}\Big) \,d \lambda  \\
&  ~+~ \Big(\coeff{1}{\sqrt{2}}\, (dt - dy) ,\coeff{1}{\sqrt{2}}\, (dt + dy) ,0,d \theta,d\varphi_1,d\varphi_2\Big)  \,,
 \end{aligned}
  \label{nullvars1}
\end{equation}
where the first vector on the right-hand side are the velocities of the radial null geodesic at $\theta =0$ with $\hat E = \frac{1}{2\sqrt{2}}$.  In particular, $r$ is to be viewed as an implicit function of $\lambda$ defined via:
\begin{equation}
 \frac{d r}{d \lambda}  ~=~  - \frac{1}{2} \, A \,\Gamma \,.
  \label{rlambdareln}
\end{equation}
Making this replacement in the metric yields:
\begin{equation}
\begin{aligned}
ds_6^2 ~=~   d\lambda \, dt  & ~-~ \frac{(r^2 + a^2)\, \Lambda}{a  R_y  A^3 \Gamma^2 } \, dt^2 ~+~ \frac{ A \, \Gamma^2\, \Lambda}{a R_y\,G } \,r^2 \, \bigg( dy +  \frac{b^2 F }{2 a^2 A^2\Gamma^2 } \,dt \bigg)^2  \\ 
  &~+~ \sqrt{Q_1 Q_5} \,\bigg[\Lambda \, d \theta^2 ~+~ \frac{1}{\Lambda} \,\sin^2 \theta\, \bigg( d \varphi_1  - \frac{1}{R_y A^2} \, dt \bigg)^2 \\
  & \qquad\qquad \qquad~+~  \frac{G}{\Lambda} \,\cos^2 \theta\, \,\bigg( d \varphi_2  - \frac{\Gamma^2}{R_y G} \, \bigg( dy +  \frac{b^2 F }{2 a^2 A^2\Gamma^2 } \,dt \bigg) \bigg)^2 \, \bigg]\,.
 \end{aligned}
  \label{transfmet}
\end{equation}
So far we have not made any approximations: this is merely a coordinate change.

For future references we note that along the locus, $\theta =0$, this metric can be regrouped into:
\begin{equation}
\begin{aligned}
ds^2  ~=~   d\lambda \, dt ~-~ & \frac{(r^2 +a^2)}{a R_y A^3 \, \Gamma^2 }  \, dt^2 ~+~\sqrt{Q_1 Q_5} \, \bigg(d \theta^2 +  \frac{r^2}{(r^2 +a^2)} d \varphi_2^2 \bigg) \\
 ~+~ &  \frac{(r^2 +a^2) \,A \, \Gamma^2}{a R_y }  \,\bigg(dy - \frac{a^2 R_y}{(r^2 +a^2)} \, d \varphi_2 +    \frac{b^2\, F }{2\, a^2 \, A^2 \, \Gamma^2}  \, d t  \bigg)^2  \,.
\end{aligned}
  \label{nullmetform}
\end{equation}
%

\subsection{The Penrose limit}
\label{ss:Penrose}

Following \cite{Blau:2002mw}, we apply the standard scaling that generates the Penrose limit:
\begin{equation}
g_{MN} ~\to~ \Omega^{-2}\,g_{MN} \,; \qquad   \lambda  ~\to~  \lambda  \,, \quad   t  ~\to~ \Omega^{2}\, t  \,, \quad  \varphi_1~\to~ \varphi_1\,, \quad   (y, \theta, \varphi_2)  ~\to~ \Omega\, (y, \theta, \varphi_2)  \,,
  \label{Penscaling}
\end{equation}
with $\Omega \to 0$.  Since we are looking at geodesics centered around $\theta =0$, the angular coordinate, $\varphi_1$  degenerates  on this locus and is therefore not  scaled like the other transverse coordinates.

The result is
\begin{equation}
\begin{aligned}
ds_6^2~=~ & d \lambda \, dt \\
 & + \sqrt{Q_1\,Q_5}\, \bigg[d\theta^2 ~+~ \theta^2\, d\varphi_1^2 ~+~ \frac{r^2}{(r^2 +a^2)}\,  d\varphi_2^2 
~+~ \Gamma^2\,  \frac{(r^2 +a^2)}{a^2\, R_y^2}\,  \Big( dy -\frac{a^2 \, R_y}{r^2 +a^2}\, d\varphi_2\Big)^2 \bigg] \\ 
~\equiv~  & d \lambda \, dt ~+~ \sum_{i,j=1}^{4} \, \cC_{ij} \, dx^i \, dx^j     \,,
\end{aligned}
\label{metPenLim1}
\end{equation}
where $\Gamma$ is given in (\ref{GamLamdefns}) and we have used (\ref{SSreg1}), (\ref{FGAdefns}) and (\ref{GamLamdefns}) to set $R_y a A =  \sqrt{Q_1Q_5}$.

In the second line of  (\ref{metPenLim1}), the transverse coordinates, $x^i$, are defined by:
\begin{equation}
x^1 ~\equiv~ \theta \, \cos \varphi_1 \,, \qquad  x^2 ~\equiv~ \theta \, \sin \varphi_1  \,, \qquad  x^3 ~\equiv~y     \,, \qquad  x^4 ~\equiv~ \varphi_2 \,,
  \label{xcoords}
\end{equation}
and the matrix, $\cC$, is given by:
\begin{equation}
\cC~\equiv~\sqrt{Q_1\,Q_5}\, \left( 
\begin{matrix}
1&0&0&0 \\ 
0&1&0&0 \\ 
0&0&  \Gamma^2\,  \frac{(r^2 +a^2)}{a^2\, R_y^2}  &- \frac{ \Gamma^2}{R_y} \\ 
0&0&- \frac{ \Gamma^2}{R_y} &\frac{r^2 + a^2 \, \Gamma^2 }{(r^2 +a^2)} \\ 
\end{matrix}
\right)  \,.
  \label{Cmat}
\end{equation}

To write this metric in a form that is better adapted to quantization of the string, we need to change coordinates to bring it to the Brinkman form:   
\begin{equation}
ds_6^2~=~2\, d  x^+ \, d  x^-  ~-~   \sum_{i,j=1}^{4} \, \cA_{ij}(x^-) \, z^i \, z^j  \, (dx^-)^2 ~+~ \sum_{i=1}^{4} \, dz^i \, dz^i  \,.
  \label{BrinkMet}
\end{equation}
Again, following \cite{Blau:2002mw}, the transformations to achieve this are
\begin{equation}
\lambda~=~2\,  x^- \,, \qquad t ~=~ x^+  ~-~  \coeff{1}{2}\, \sum_{i,j=1}^{4} \, \cM_{ij}(x^-) \, z^i \, z^j  \,, \qquad x^i ~=~  \sum_{j=1}^{4} \, \cQ^i{}_{j}(x^-) \,   z^j   \,,
  \label{coordtrf1}
\end{equation}
where $\cQ^i{}_{j}$ and $\cM_{ij}$ are defined via:
\begin{equation}
\cC_{ij}\,\cQ^i{}_{k}\,\cQ^j{}_{\ell}  ~=~  \delta_{k\ell}\,,\qquad \cC_{ij}\,\big( \cQ'{}^i{}_{k}\,\cQ^j{}_{\ell}  - \cQ'{}^i{}_{\ell}\,\cQ^j{}_{k} \big)~=~  0 \qquad \cM_{k \ell} ~=~ \cM_{ \ell k} ~=~  \cC_{ij}\,  \cQ'{}^i{}_{k}\,\cQ^j{}_{\ell} \,,
  \label{QMdefns}
\end{equation}
where $'$ denotes $\frac{d}{dx^-}$.  The first equation fixes $\cQ$ up to an orthogonal transformation, and this freedom is then fixed by the second identity, which, in turn, guarantees the symmetry of $\cM$.  The matrix, $\cA$, in  (\ref{BrinkMet}) is then given by:
\begin{equation}
\cA_{ij} ~=~ + \bigg[\frac{d}{dx^-} \Big(\cC_{k\ell}\, \cQ'{}^\ell{}_{j}\,\Big) \bigg] \,\cQ^k{}_{i}\,.
  \label{cAres}
\end{equation}
One should note our matrix, $\cA_{ij}$, defined in (\ref{BrinkMet}) and (\ref{cAres}), has the opposite sign to the conventional form in  \cite{Blau:2002mw}.  We make this choice because,   in Section \ref{sec:stringprop},  we show that $\cA_{ij}$ defines a mass matrix for the  modes of a string probe.

The Brinkman transformation is trivial on 
\begin{equation}
(z^1, z^2) ~=~  (x^1, x^2) ~=~ (Q_1\,Q_5)^\frac{1}{4} \,\theta \, (\cos \varphi_1  \,, \sin \varphi_1) \,,
  \label{trivtrf}
\end{equation}
and non-trivial on $(x^3, x^4) = ( y\,,  \varphi_2 )$.  We therefore specialize to this two dimensional space and the transformation from $(x^3, x^4)$ to   $(z^3,z^4)$.  We find:
\begin{equation}
\cQ ~=~ \cQ_0 \, \cR \,, \qquad  
\cQ_0 ~\equiv~ \sqrt{\frac{a\, R_y}{A\,(r^2 +a^2)}} \, \left(\begin{matrix}
\frac{1}{\Gamma}&\frac{a}{r} \\ 
0&\frac{(r^2 + a^2)}{a\, R_y\, r} 
\end{matrix} \right) \,, \qquad
\cR ~\equiv~  \left(\begin{matrix}
\cos \eta(r) & -\sin \eta(r)  \\ 
\sin \eta(r)&\cos \eta(r) 
\end{matrix} \right) \,,
 \label{QRmats1}
\end{equation}
The middle equation in   (\ref{QMdefns}) then yields 
\begin{equation}
\frac{d}{d r} \, \eta(r)  ~=~ - \frac{a\, \Gamma}{(r^2 +a^2)} \,.
  \label{etaeqn}
\end{equation}
The non-trivial $2 \times 2$ part of $\cA_{ij}$ acting on $(z^3,z^4)$ is then given by:
\begin{equation}
\cA ~=~ \cR^{-1} \left(\begin{matrix}
\mu_0 + \mu_1 & \mu_2  \\ 
\mu_2 &\mu_0 - \mu_1 
\end{matrix} \right)    \, \cR \,, 
 \label{Aform1}
\end{equation}
where
\begin{equation}
\begin{aligned}
\mu_0 ~=~ &  \frac{b^2 \, r^{2n}}{2\,(r^2 +a^2)^{n+2}}\, \bigg[\Gamma^2 ~+~  \frac{n^2 \, a^2}{r^2}   \bigg]\,,  \\ 
\mu_1 ~=~ & -  \frac{b^2 \,  r^{2n}}{(r^2 +a^2)^{n+2}}\, \bigg[ \Gamma^2 ~-~  \frac{n(n-1) \, a^2}{2\,r^2}   \bigg] \,, \\ 
\mu_2 ~=~ & \frac{2\, n\, a\, b^2 \,r^{2n-1}}{(r^2 +a^2)^{n+2}}\,  \Gamma \,.
\end{aligned}
 \label{mures1}
\end{equation}

Note that for global AdS$_3 \times S^3$ one takes $b=0$ and has:
\begin{equation}
\cA_{ij}   ~\equiv~ 0 \,, 
 \label{Avan}
\end{equation}
which means that the Penrose limit is simply flat space.   Similarly,  the BTZ limit is given by taking $a \to 0$, and in this limit one finds  $\Gamma \to 0$ and that $\cA$ vanishes identically.  Once again, the Penrose limit  reduces to flat space.  Thus non-triviality of the Penrose limit requires $a, b \ne 0$. 

\begin{figure}[h!]
\centering
\includegraphics[width=.75\textwidth]{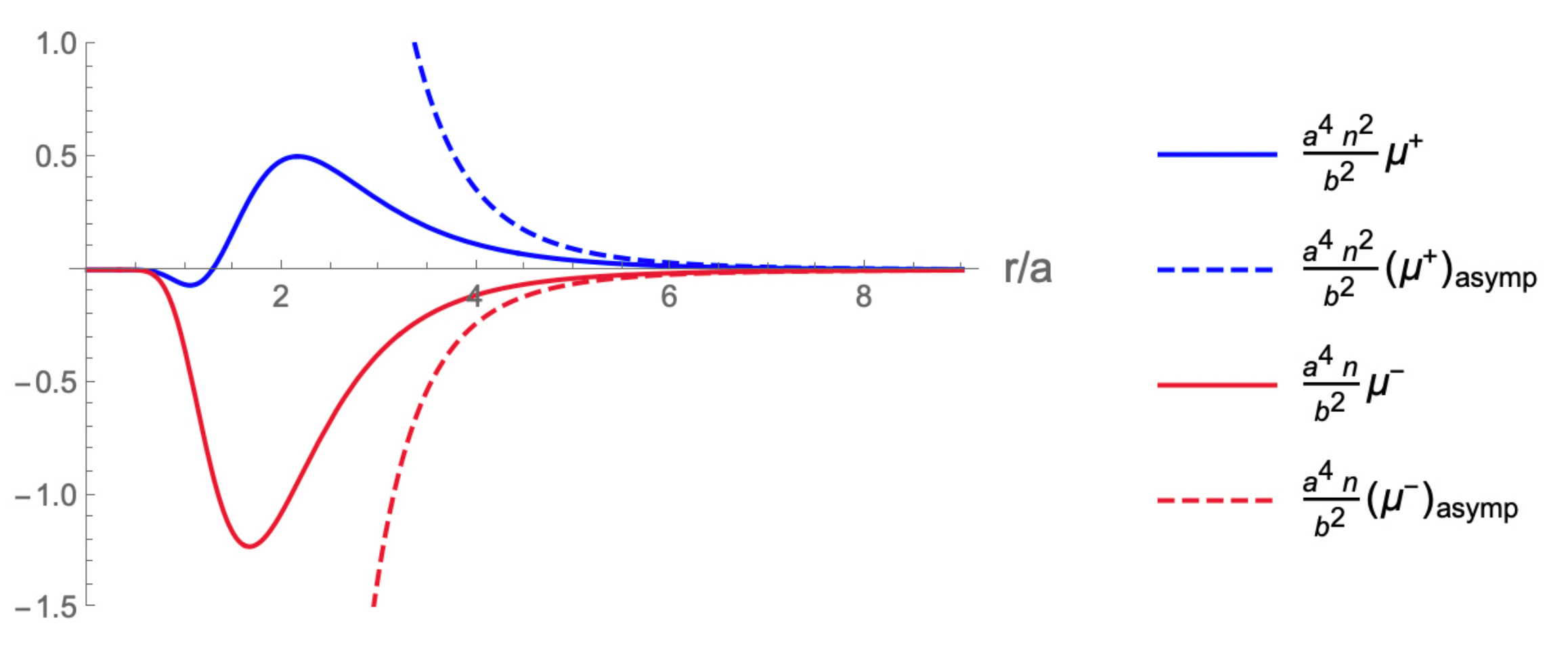}
\caption{\it Eigenvalues of $\cA$ in the $z^3$-$z^4$ plane, and their approximation by their leading power law behavior for $a\ll r\ll b$, for $n=9$. Note that for these parameters, the edge of the cap lies at $\frac{r}{a} \sim 3$.}
\label{fig:bumpfns}
\end{figure}

The eigenvalues of $\cA$  drive the behavior of the string because they generate  the effective masses, and a positive mass compresses and stabilizes the string while a negative mass produces an instability that excites the string.  Since we are going to work in the limit in which $a \ll r,b$, it is instructive to investigate this at small $a$.  From  (\ref{mures1}) one has:
\begin{equation}
\begin{aligned} 
\mu_0 &~\sim~  \frac{b^2}{2r^4}\Bigl[\Gamma^2+\frac{n^2a^2}{r^2}\Bigr] ~\sim~   \frac{a^2b^2n(n+1)}{2r^6} \,, 
\\[.2cm]
\mu_1 & ~\sim~ - \frac{b^2}{r^4}\Bigl[\Gamma^2-\frac{n(n-1)a^2}{2r^2}\Bigr] ~\sim~   \frac{a^2b^2n(n-3)}{2r^6} \,, 
\\[.2cm]
\mu_2 &~\sim~    \frac{2nab^2\Gamma}{r^5} ~\sim~   \frac{2\, n^{\frac32} a^2b^2}{r^6}  \,, 
\end{aligned}
\label{muexpand}
\end{equation}
which leads to the eigenvalues of $\cA$:
\begin{equation}
\mu^\pm ~=~ \mu_0 ~\pm~ \sqrt{\mu_1^2 + \mu_2^2} ~\sim~ \frac{a^2b^2}{2 \, r^6}\, n \, \Big[\, (n+1) ~ \pm~ \sqrt{(n+1)(n+9)}  \, \Big]  \,.
 \label{evals1}
\end{equation}
This shows that there is a positive mass that grows with $n^2$ and a negative mass that grows linearly in $n$.  It is the latter that will be of most interest.  A plot of $\mu^\pm$ and their scaling limits in the region $a\ll r\ll b$ are shown in Figure~\ref{fig:bumpfns}.
%
\vspace{0.5mm}
\vspace{1mm}

Before concluding this section, we note that (\ref{Aform1}) implies that the eigenvectors of $\cA_{ij}$ rotate through the action of $\cR$ as the string passes through the geometry.  We now show that this rotation is very small in the throat and bounded by $\frac{\pi}{2}$ in the range $0 < r < \infty$.   First observe that  since $\Gamma \le 1$, (\ref{etaeqn}) implies
\begin{equation}
\frac{d}{d r} \, \eta(r)  ~\le~ - \frac{a}{(r^2 +a^2)}  \qquad \Rightarrow \qquad  | \eta(r)|  ~\le~ \Big | \arctan \frac{r}{a} \, \Big| \,, 
 \label{etabound}
\end{equation}
which establishes the overall bound.  Indeed, outside the cap, one has $a \sqrt{n} < r < \infty$ and 
\begin{equation}
\delta | \eta(r) |  ~\le~ \Big | \frac{\pi}{2} - \arctan  \sqrt{n} \, \Big|  \,.
 \label{etabound2}
\end{equation}
This means that the rotation outside the cap is less than $\frac{\pi}{4}$,  and becomes smaller with increasing $n$.  Thus the unstable, negative-mass direction remains within a narrow sector of the $(y,\varphi_2)$-plane. 

The direction  of stretching forces is the eigenvector associated to $\mu^-$ and this direction adiabatically rotates as the string travels down the throat.  Because the metric direction that is stretching shifts from being entirely along $y$ asymptotically to being an equal admixture of $y$ and $\varphi_2$ by the time one reaches the cap, as one sees from the last line of~\eqref{nullmetform}~-- the stretching comes from the momentum carried by the background, associated to the $dt\,dy$ and $dt\,d\varphi_2$ terms in the metric.

\subsection{The string metric}
\label{ss:stringmet}

In order to probe the background using  strings, we should use the ten-dimensional string metric, and to describe this, we need to introduce the electrostatic potentials underlying the superstratum solution. These potentials are given by:
\begin{equation}
\begin{aligned}
Z_1 ~=~ &  \frac{Q_1}{\Sigma} + \frac{R_y^2 \, b^2 }{2 \, Q_5\, \Sigma} \, \Delta_{2,0,2n} \, \cos \chi_{2, 0,2n}   \,,  \qquad Z_2  ~=~\frac{Q_5}{\Sigma} \,, \\ 
Z_4  ~=~  & \frac{b\,   R_y}{\Sigma} \, \Delta_{1,0,n}   \;\! \cos \chi_{1,0,n} \,,   \qquad \qquad \cP ~\equiv~  Z_1 \, Z_2  -  Z_4^2 \,, 
\end{aligned}
\label{Z1Z2Z4}
\end{equation}
where
\begin{equation}
\begin{aligned}
\chi_{k,m,n} & ~\equiv~  \coeff{\sqrt{2}}{ R_y}\, (m+n) \, v ~+~  (k-m) \, \varphi_1 ~-~   m \,  \varphi_2 \,,  \\
 \Delta_{k,m,n} &~\equiv~ \frac{a^k \, r^n }{(r^2+a^2)^{\frac{k+n}{2} }}\,\sin^{k-m}\theta\,\cos^m\theta \,, \qquad \Sigma ~\equiv~ r^2 + a^2 \cos^2\theta 
 \end{aligned}
 \label{Deltachidefns}
\end{equation}
Note that in the backgrounds we consider this  paper we have to $k=1,m=0$ but  $n$ is arbitrary.  Also note  that $\cP$ is given by
\begin{equation}
\cP~=~   \frac{Q_1Q_5}{\Sigma^2} \, \Lambda^2 \,,
\label{cPsimp}
\end{equation}
and is independent of the oscillations in $\chi_{k,m,n}$.

The string metric is then given by \cite{Giusto:2013rxa,Bena:2015bea,Bena:2017xbt}. 
\begin{equation}
d s^2_{10} ~=~\sqrt{ \frac{Z_1 Z_2}{\cP}} \,  \,ds^2_6 ~+~ \sqrt{\frac{Z_1}{Z_2}}\,d \hat{s}^2_{4}~=~\sqrt{ \frac{Z_1 Z_2}{\cP}} \,  \bigg( \,ds^2_6 ~+~ \frac{\sqrt{\cP}}{Z_2}\,d \hat{s}^2_{4} \bigg) \,,
\label{tenmetric}
\end{equation}
where $ds^2_6$ is the metric analyzed in the previous sections and $d \hat{s}^2_{4}$ is the trivial metric on the $\IT^4$ of the  compactification.  

For our analysis, the important thing to note is that the oscillations only arise in the metric warp factors through $Z_1$ or $Z_4^2$, and so always come with a factor of $\sin^2 \theta$.   Moreover, expanding around $\theta =0$, we have  
\begin{equation}
\sqrt{ \frac{Z_1 Z_2}{\cP}}  = 1 + (1-G)\, \theta^2 \, \cos^2 \chi_{1,0,n}   
~~,~~~~ 
\sqrt{ \frac{Z_1}{ Z_2}}  = \sqrt{ \frac{Q_1}{ Q_5} }\, \Big( 1 + (1-G)\, \theta^2 \, \cos \chi^{~}_{2,0,2n} \Big) \,,
\label{warpexp}
\end{equation}
where $G$ is defined in (\ref{FGAdefns}).  
More generally, expanding around some other radially directed null geodesic at a generic point on $\IS^3$ leads to a warp factor in the metric that scales as
\be
\sqrt{ \frac{Z_1 Z_2}{\cP}}  \sim 1+\frac{a^2}{r^2} \cos^2\!\chi^{~}_{1,0,n} \,\sin^2\theta
\ee
in the region of large tidal forces $a\ll r\ll b$.  Compared to the leading tidal effects, the contributions of the warp factor relating string frame to Einstein frame are down by a factor of $a^2/r^2$, and so can be  neglected.%

One consequence of these properties of the warp factors relating string and Einstein frames is that the geodesics along $\theta =0$ in the string metric, (\ref{tenmetric}), are the  same as in the trivial combination of $ds^2_6$ and $d \hat{s}^2_{4}$.  Thus for the geodesics along $\theta =0$ we can work entirely with $ds^2_6$.   
Next, recall that the Penrose limit involves a quadratic expansion around the null geodesics along $\theta =0$.  We note  that  (\ref{metPenLim1}) is the leading quadratic expansion of  (\ref{transfmet})  under the scaling (\ref{Penscaling}). Adding the warp factor $\sqrt{ \frac{Z_1 Z_2}{\cP}}$ will simply multiply the result by $1$, and the $\theta^2$ terms in  (\ref{warpexp}) will only contribute at fourth order in the scaling transformation (\ref{Penscaling}).
Finally, observe that the Penrose limit of the torus directions will involve scaling the torus coordinates by $\Omega$.  The $\theta^2$ terms in $\sqrt{ \frac{Z_1}{ Z_2}}$ will thus be of fourth order and will not contribute in the Penrose limit.  As a result, the torus directions will only contribute four uninteresting flat directions to the Penrose limit.   

We therefore conclude that the six-dimensional metric,  $ds^2_6$,  is all that is required to capture the essential physics of the full string metric so long as one focusses on the geodesics with $\theta =0$, and on the corresponding Penrose limit of the metric in the neighborhood of these geodesics.  We also expect that $\theta=0$ is not special in this regard, given that the contributions of the warp factor at other values of $\theta$ are sub-leading in $a/r$.

\section{String propagation}
\label{sec:stringprop}

The propagation of strings in $(3+1)$-dimensional, plane-fronted wave backgrounds was analyzed in~\cite{Horowitz:1990sr}, and the generalization to higher dimensions is straightforward.  This class of backgrounds has a Brinkman form~\eqref{BrinkMet}, \eqref{Blim2}.
 In this background, the equations of motion of the worldsheet string theory simplify and, in the light-cone gauge $x^-=\alpha'E\tau$,  the equations of motion of the transverse coordinates reduce to
a set of free fields coupled via a time-dependent mass matrix.
As is usual in string theory, we parametrize the world-sheet in terms of dimensionless parameters, $(\sigma,\tau)$.  This means that $E$ has dimensions of mass, or inverse length. 

Here we have a $(9+1)$-dimensional  metric, but in the Penrose limit the two transverse coordinates $x^1$ and $x^2$, as well as the $\IT^4$ directions, are, to a good approximation, simply massless free fields and whose states decouple from the metric.  The remaining coordinates, $x^3$ and $x^4$, have  a non-trivial mass matrix, and they also couple to $x^1$ and $x^2$  through the $B$-field, however we will show that the latter effect is sub-leading and can be neglected.  For the superstratum, the quantities $\mu_n$ in the matrix $\cC$ that determines the plane wave profile in $x^3$ and $x^4$ are bump functions that die off as $r^{-6}$ at large $r$; they peak in the vicinity of  $r\sim a\sqrt{n}$ and then (for $n>1$) fall smoothly to zero around $r\sim a$ and vanish in the region $0<r< a$.    We work in the deep throat limit $a\sqrt{n} \ll r < b$.
In~ \cite{Tyukov:2017uig,Bena:2018mpb}, it was shown that tidal forces grow large at a radial position midway down the throat, namely $r\sim\sqrt{ab}$.

\subsection{Tidal excitation of  strings}
\label{ss:stringexcitation}

The bosonic part of the string worldsheet action with the Brinkman form of the NS fields~\eqref{BrinkMet}, \eqref{Blim2} becomes;
\begin{align}
\label{WSaction}
\cS &= \int d\tau \, d \sigma\, \Bigl[ \partial_a x^+ \partial^a x^- + \partial_a z^i \partial^a z^j  \delta_{ij}
\nn\\ &\hskip 2cm  
- \cA_{ij}(X^-) \,z^i z^j \,\partial_a x^- \partial^a x^-
+ \cB_{- i}(X^-,z)  \,\epsilon^{ab}\partial_a x^- \partial_b z^i \Bigr]
\end{align}
where $\theta$ is the radial coordinate in the $z^1$-$z^2$ plane,~\eqref{trivtrf}.  In the light-cone gauge
\be
x^- = \alpha' E\tau
\label{lightconemom}
\ee
the metric term $\cA_{ij} z^iz^j$ is an effective time-dependent mass matrix that is trivial in the $z^1$-$z^2$ plane and non-trivial in the $z^3$-$z^4$ plane, with one positive and one negative eigenvalue.  The B-field is also quadratic in the transverse coordinates, and acts to rotate the string worldsheet in the $z^1$-$z^3$ plane and in the $z^2$-$z^4$ plane.   For $a\ll r<b$, one can use (\ref{muexpand}) and (\ref{evals1}) to determine the eigensystem of (\ref{Aform1}):%
\footnote{Note that $z^\pm$ here are not some rescaling of the longitudinal coordinates $x^\pm$, but rather the eigendirections in the $x^3$-$x^4$ plane of the tidal tensor.}
\eqn{
\label{mu evalues}
z^\pm = \Big(\coeff{n-3\pm\sqrt{(n-1)(n-9)}}{4\sqrt{n}}\, ,\, 1\Big)  ~~&,~~~~ 
\mu_\pm = \frac{a^2b^2}{2r^6} \, n\Big(n+1\pm\sqrt{(n+1)(n+9)}\,\Big) ~.
}
As mentioned above, the unstable direction adiabatically rotates in the $y$-$\varphi_2$ plane as the string travels down the throat.
The rotation angle $\eta$ in~\eqref{QRmats1} does not change much in the region of large tidal forces (it is between zero and $\pi/4$ in this region), and in what follows we ignore its effects.
Note that along the null geodesic we are expanding around, one has
\be
\label{x-to-r}
x^- \approx - \frac{r^2}{b\sqrt{2n}} ~.
\ee
For the $B$-field one has
\be
\cB_{-3}\sim \frac{b\,\Gamma}{r^2}\,z^1 \sim \frac{abn^{\frac12}}{r^3}\,z^1 \equiv B_{31} z^1
~~,~~~~
\cB_{-4} \sim \frac{ b\,\Gamma}{r^2}\,z^2 \sim \frac{abn^{\frac12}}{r^3}\,z^2 \equiv B_{42} z^2 ~.
\ee
The effect of the $B$-field is suppressed relative to the mass matrix; it will not dramatically alter the behavior of the unstable mode, but rather just shift slightly the unstable mode into a mixture of the $z^-$ and a similar linear combination of the $z^1,z^2$ directions; again we will ignore its effects.
In Penrose limits, the string action in Green-Schwarz variables is also quadratic in fermions~\cite{Russo:2002qj,Mizoguchi:2002qy}.  While we haven't examined the effects of the RR antisymmetric tensor fields on the string worldsheet dynamics, we expect that in a light-cone Green-Schwarz formalism one will have an $x^-$ dependent mass term.
In the throat region $a\ll r\ll b$, both the $B$-field and the metric exhibit a characteristic deviation from the BTZ geometry in the leading multipole.  We expect this behavior extends to the RR antisymmetric tensors of the superstratum as well, leading to an effective time-dependent mass for worldsheet fermions.  This time-dependent mass should lead to production of fermion modes, but like the effects of the $B$-field, we suspect that this is a subleading effect, though we have not examined the details.  Regardless, the production of worldsheet fermion excitations merely robs the string of more of its initial kinetic energy, diverting it into additional string oscillator excitations.

Variation of the worldsheet action~\eqref{WSaction} yields the equations of motion for the $k^{\rm th}$ oscillator modes
\be
\label{modeeq}
\partial_{\tauhat}^2 z^i + {k^2} \,z^i + {(\alpha'E)^2}\cA_{ij}(\tauhat) z^j + i{k}\,{\alpha'E}\, B_{ij}(\tauhat)z^j = 0
\ee
which has the form of a system of coupled harmonic oscillators.  The interesting coordinate is the unstable mode $z^-$, which is stable for large $r$ but then goes unstable at a transition point $r\approx r_*$ midway down the throat (in the sense of the geometric mean), 
\be
\label{rstar}
{k^2} = \frac{2\,a^2b^2n{(\alpha'E)^2}}{r_*^6} \approx -\frac{2\,a^2}{b\sqrt{n}(\alpha'E) \tauhat_*^3} ~,
\ee
where we have used (\ref{x-to-r}) to replace the radial coordinate $r$ by $\tauhat=\frac{x^-}{\alpha'E}$.  At the value  of  $r$, or $\tauhat$, defined by this  equality, the natural oscillator frequency is overwhelmed by tidal forces and the string starts to stretch exponentially in the $z^-$ direction in its $k^\th$ mode.  The larger $k$ is, the further down the throat one must go before the instability sets in.  

Overall, this system of modes is qualitatively similar to the spectrum of fluctuations in multifield inflation~-- a system of harmonic oscillators, with one unstable mode, coupled together (see~\cite{Wands:2007bd,Langlois:2010xc} for reviews)\footnote{For an embedding of inflationary physics into string worldsheet dynamics, see~\cite{DaCunha:2003fm,Martinec:2014uva}.}.
There, the inflaton traveling down the potential energy landscape in general travels a curved trajectory in field space, and there is mixing of the various fluctuation modes, \ie\ the dominant unstable direction and the directions locally transverse to it.  
%
The effective magnetic field and the field space rotation parametrized by $\eta$ also act to divert some of the $z^-$ runaway's kinetic energy into other directions, but these effects are a sub-leading effect here.  
The local eigenvectors $z^\pm$ in equation~\eqref{mu evalues} have an approximately fixed orientation in the region of large tidal forces.  As a result, at leading order we can simply concentrate on the single unstable mode and ignore the rest.

In this analogy to the analysis of quantum fluctuations during inflation, the time $\tauhat_*(k)$ is the analogue of the horizon crossing time for a given fluctuation mode of the inflaton, where the mode starts to grow exponentially.  Eventually one reaches the bottom of the throat at $r\sim a\sqrt{n}$, where the geometry transitions to a smooth $AdS_3$ cap, which occurs at a time 
\be
\tauf = -\frac{a^2\sqrt{n}}{\sqrt{2}\,b{\,\alpha'E}} 
\ee
along the string trajectory.  This transition is analogous to the end of inflation, where the growth of modes shuts off.  At this point  the affected string modes can start oscillating again.  Between $\tauhat_*(k)$ and $\tauf$, the $k^\th$ mode is growing exponentially by a factor which we now estimate.

We can use a WKB approximation for evolution of the modes.  
Focusing on a single unstable mode, the WKB integral for $z^-$ is 
\begin{align}
\label{WKBgrowth}
z^- &\approx\; 
\exp \Big[-i\int^{\tauhat} \!\!d\tauhat' \, \omega(\tauhat') \Big] =
\exp \Big[-i\int^{\tauhat} \!\!d\tauhat'\Big( k^2 - \coeff{a^2}{b\sqrt{2n} (\alpha'E) (\tauhat')^3}\Big)^\half\, \Big]
\nn\\[.3cm]
&= 
\exp\Big[\Big(\coeff{2\,a}{r}\Big) {}_2F_1\Big(\tight-\hf,-\coeff16,\coeff56 ; \coeff{k^2\,r^6}{2n(\alpha'E a b)^2 } \Big) \Big]
\nn\\[.3cm]
& \sim \begin{cases}
\exp\Big[-i \coeff{k}{\alpha'E} \coeff{r^2}{b\sqrt{2n}} \Big]  \quad,~~~~ r_*(k)< r<b \\[.2cm]
\exp\Big[{{2}}\,a\Bigl( r^{-1} - r^{-1}_*(k)\Big)\Big] \qquad,~~~~ a\sqrt{n}<r<r_*(k) 
\end{cases}
\end{align}
where $r_*(k)$ is given in~\eqref{rstar}. 
The WKB approximation is valid for early and late times, and only fails in a small region around the onset of instability $\tauhat_*$.  The early time solution is a plane wave with an adiabatically evolving frequency; the late time solution as $\tauhat\to 0^-$ is given in terms of Bessel functions. 
The form of the late time solution sets in rather quickly, as the hypergeometric function is well-approximated by a constant (unity) when its argument is between zero and one.
The total mode amplification in the unstable region is of order one because, while the tidal forces exceed the string scale rather far up the throat, the time from this point to the bottom of the throat is correspondingly small; the net effect is that the total impulse provided to the string modes remains small by the time the string reaches the cap.  In other words, while the physics in this region is one of inflation of the fluctuation modes, the number of e-folds of inflation is of order one, and so mode production is of order one for the modes that go unstable in the throat, rather than being exponentially large.  Remarkably, while the tidal stress grows with the incident energy $E$, the time interval over which these forces act shrinks with $E$, with the overall net energy provided to the mode being independent of $E$.

We estimate the mode production as the probe string passes through the geometry in stages.  The string evolves as a plane wave with an adiabatically evolving frequency until it reaches the vicinity of the instability of the lowest oscillator mode at $r=r_*(1)$.  From this point until the beginning of the cap at $r\sim a\sqrt{n}$, one by one the low modes of the string go unstable.  For $n>1$, the instability shuts off\,\footnote{For $n=1$ the tidal forces remain constant across the cap, but the proper time  it takes to cross the cap is very  small, and so the tidal impulse also remains modest.}   as the string enters the cap region where the geometry is approximately global $AdS_3\times \IS^3$.  The string cruises through this region until it exits on the other side of the cap, where it starts moving back up the throat.  Once again it is in the region of mode instability, and so the unstable modes grow again until one by one they reach their transition point $r_*(k)$ and become stable again.  Above $r\sim r_*(1)$ all modes of the string go stable again, but its overall state is a superposition of various mass eigenstates.

In the region where the mode production is taking place, the string is relativistic and so the Penrose limit is a reasonable approximation to the geometry.  We can use the Penrose limit to estimate the mode production, but then return to the full geometry to estimate the turning point of the excited string's motion where it reaches ``apostratum'' and then falls back down toward the cap.  The reason we need to make this switch is that the Penrose limit (where the light-cone analysis of the string dynamics holds) only applies to relativistic motion of the string; on the other hand, to find the turning point we are obviously in the non-relativistic regime.

The WKB estimate of the Bogoliubov coefficient for modes that go unstable as the string passes down the throat to the cap is thus
\be
|\beta_k|^2 \approx \exp\left[ 4 \left(n^{-\frac12}-\Big(\frac{a^2k}{\sqrt{n}\,b\,\alpha'  E}\Big)^{\frac13}\right)\right] ~.
\ee
Mode production effectively shuts off when $r_*(k_{\rm max}) \sim a\sqrt{n}$, \ie
\be
\label{kmax}
k_{\rm max} \sim \frac{b\,\alpha'  E}{a^2n}
\ee
because there is no region of the throat or cap where modes with $k>k_{\rm max}$ go unstable.  However for a long throat $b\gg a$ and sufficiently large energy $E$, there will be many modes that do experience a regime of instability.  

For modes with $k\gg k_{\rm max}$, mode production is exponentially suppressed.  One can again use WKB methods to estimate the Bogoliubov coefficient $\beta_k$ %
\cite{Audretsch:1979uv,Lawrence:1995ct,Silverstein:2014yza}.  
The wave equation~\eqref{modeeq} is equivalent to the time-independent Schr\"odinger equation for a quantum mechanical particle with energy eigenvalue $k^2$ and potential given by the bump function $\mu_-(r)$.  
Consider a pure left-moving wave incident from the right (say) in the initial state, and a combination of $R$ times a right-moving wave and $T$ times a left-moving wave in the final state.  One has a superposition of left- and right-moving waves at $x\to +\infty$ and only a right-moving wave at $x\to -\infty$.
Normalizing the incident wave to unit amplitude, the standard reflection and transmission coefficients $R,T$ satisfy
\be
|R(k)|^2+|T(k)|^2=1 ~.
\ee
In the current application to the Klein-Gordon equation with a time-dependent mass, one has a pure positive frequency wave in the far past, and a superposition of positive and negative frequency waves in the far future, with amplitudes $\alpha$ and $\beta$ respectively.  So space in the QM problem becomes time in the mode creation problem with the far past $t\to -\infty$ mapped to $x\to -\infty$ and $t\to+\infty$ mapped to $x\to +\infty$.
Conservation of the KG norm is $|\alpha|^2-|\beta|^2=1$.  Comparing to the QM problem, the transmitted wave there is the initial state wave here, so we have rescaled things by a factor of $T$ to give the wave at $x\to -\infty$ unit amplitude, and as a consequence $\alpha=1/T$ and $\beta=R/T$.
The regime where $\beta$ is exponentially suppressed corresponds to the amplitude for above-barrier reflection in the associated quantum mechanics problem.   

The dynamics in the full bump function potential $\mu_-(r)$ is difficult to evaluate in closed form, even in the WKB approximation.  Instead we consider an alternative bump function potential
\be
V_{\rm bump} = V_0\, {\rm sech}^2[\kappa(\tau-\tau_0)]
\ee
having approximately the same height and width, in order to elucidate the physics.
The reflection coefficient for this potential is known~\cite{Eckart:1930zza}
\be
\label{betabump}
\beta_k \approx \exp\big[-\big(k-\sqrt{V_0}\,\big)/\kappa\big] ~.
\ee
The parameters in the potential approximating $\mu_-(\tau)$ are
\be
V_0 \approx \frac{(b\,\alpha'  E)^2}{2a^4n^2} \approx k_{\rm max}^2
~~,~~~~
\kappa \approx \frac{b\,\alpha'  E\sqrt{2n}}{a^2}
~~,~~~~
\tau_0 \approx \frac{a^2}{2b\,\alpha'  E}
\ee
and so the amplitude to excite modes with $k>k_{\rm max}$ is suppressed exponentially in $\frac{(k-k_{\rm max})a^2}{b\,\alpha'  E\sqrt{2n}}$.

One can understand this result and its genericity in the WKB approximation.  Ordinarily in the WKB method, one integrates the action differential $pdx$ along the classical trajectory to determine the phase of the semiclassical wavefunction; here however, the classical trajectory corresponds to the transmitted wave because the incident energy is above the top of the barrier.  We are instead interested in the exponentially small part of the amplitude for the reflected wave.  This is obtained by departing the real axis in $x$ to find the closest turning point $x_0$ in the complex plane, where $p=\sqrt{k^2-\mu_-(\tau)}$ has a branch point.  Sending the integration contour around the branch singularity, the momentum returns to the real axis with the opposite sign.  The wavenumber changes from $k-k_{\rm max}$ to zero in traversing this contour, over a position space region of width $\delta\tau\sim 1/\kappa$, and so the WKB integral will take the value $(k-k_{\rm max})/\kappa$ up to a constant of order one, thus explaining why the result~\eqref{betabump} is a reasonable approximation for the Bogoliubov coefficient in the regime $k>k_{\rm max}$.

We can now estimate the average mass of the string as it emerges from the cap.  The Bogoliubov coefficients are slowly varying over the range $k<k_{\rm max}$, and then exponentially decay with a characteristic falloff set by $\kappa\sim k_{\rm max} n^{3/2}$.
Approximating the mode sum by an integral, we can estimate the average energy of the final state string
\be
\label{mass estimate}
\big\langle \alpha' m^2 \big\rangle ~=~  \langle \cN_{\rm osc} \rangle ~\equiv~   \int {dk}\, k|\beta_k|^2   ~\sim~   \textit{const.}\, k_{\rm max}^2 
~\sim~ \textit{const.}\Big(\frac{b\, \alpha' E}{n\,a^2}\Big)^2 
\ee
with the constant of order one.

Because the string is a superposition of various mass eigenstates, those mass eigenstates will spatially separate, as different mass strings of a given energy $E$ will reach different turning points of their radial motion where they attain their maximum radius and then fall back toward the cap.  Eventually the string will settle into the cap after enough passes up and down the throat.

\subsection{Trapping of the string}
\label{ss:trapping}

Our goal now is to estimate the height to which the excited string returns after its strong tidal interaction with the superstratum geometry.  As discussed above, we approximate the dynamics in stages.  For the initial infall, we assume that the string is relativistic as it enters the region of large tidal forces; we can either take it to be a massless supergraviton state, or a massive string that has been dropped from a sufficient height that its blueshift in the throat makes it highly relativistic.  We then approximate the geometry in the Penrose limit appropriate to the expansion around the nearly null geodesic being followed by the probe center-of-mass.  The string is tidally stretched in the $y$ and $\varphi_2$ directions until it reaches the cap; as we will see, it is still relativistic at this point, so the Penrose approximation continues to hold.  The string will then pass almost ballistically through the cap in the region $r\lesssim \sqrt{n} \, a$, where the tidal forces are negligible.  At this point, the string starts to head back up the throat and into the region of large tidal forces.

For simplicity, we assume that the center of mass motion continues along a purely radial trajectory, moving only in the $(t,r)$, or $(t,\lambda)$, plane.   It is possible that the interactions with the background in the region of large tidal stresses could deflect the string into some of the angular directions, however this would take kinetic energy out of the radial motion and result in an even lower return height.

We now analyze the  ``coasting phase,''  in which the string is no longer undergoing  strong tidal forces, and determine the return height of the string.  We assume that when  the string enters this phase, it is still moving at relativistic speeds and has not spread sufficiently to invalidate the use of  the Penrose limit.  Under these  circumstances, the string is well described by motion in the metric (\ref{BrinkMet}), but now the dynamical mass, $\cA_{ij}$, is negligible.  The Virasoro constraint obtained from pulling back the geometry to the string worldsheet is
\begin{equation}
\begin{aligned}
0  ~=~&  \frac{1}{2\pi} \, \int_{0}^{2\pi} \,  \big(G_{MN}+B_{MN}\big)\Big(\partial_\tau x^M\partial_\tau x^N + \partial_\sigma x^M\partial_\sigma x^N \Big) \, d\sigma \\
 ~\approx~ & \bigg(  \frac{ds}{d\tau} \bigg)^2_{\rm c.o.m.}  ~+~ \frac{1}{2\pi} \, \int_{0}^{2\pi} \,   \Big(\big( \partial_\tau z^i \big)^2 + \big( \partial_\sigma z^i \big)^2\Big)   \, d\sigma\,,
\end{aligned}
  \label{Vir1}
\end{equation}
where we have restored the center of mass dynamics and written it as the geodesic action.  The transverse oscillations contribute to the constraint through their mean-square.   This is  simply $\alpha'$ times the expectation of the number operator, $\langle \cN_{\rm osc} \rangle$, for the excited string, and is dominated by the unstable direction.  The term, $\alpha'\, \langle \cN_{\rm osc} \rangle$, is thus playing the role of a mass parameter for the time-like geodesic.  Once again, we are  ignoring the effects of worldsheet fermions. (Production of fermion modes will only further rob the string of its center of mass kinetc energy).

Since the string is  massive,  we will restore the terms we dropped in the Penrose approximation and use the full metric~\eqref{nullmetform}.  In particular, as the string slows down, the term proportional to $dt^2$ can no longer be neglected.

We can now read off the action for this geodesic from (\ref{nullmetform}): 
\begin{equation}
0  ~=~ \bigg( \frac{d t}{d \tau}\bigg) \bigg( \frac{d \lambda}{d \tau}\bigg)  ~-~  \frac{(r^2 +a^2)}{a R_y A^3 \, \Gamma^2 }  \,  \bigg( \frac{d t}{d \tau}\bigg)^2 ~+~  \alpha'\, \langle \cN_{\rm osc} \rangle\,,
  \label{Vir3}
\end{equation}
where  we have frozen the angular motions of the center of mass.   The conserved energy associated with the Killing vector $\frac{\partial}{\partial t}$ is: 
\begin{equation}
\frac{1}{2}\, \bigg( \frac{d \lambda}{d \tau}\bigg)  ~-~  \frac{(r^2 +a^2)}{a R_y A^3 \, \Gamma^2 }  \,  \bigg( \frac{d t}{d \tau}\bigg)  ~=~  \alpha' E \,.
  \label{Energy1}
\end{equation}
%
%
Note that in the relativistic regime, the second term can be neglected and one recovers the usual light-cone gauge expression: 
\begin{equation}
x^-  ~=~  \frac{1}{2}\, \lambda ~=~  \alpha' E \, \tau \,.
  \label{lcgauge}
\end{equation}

Using the energy integral, (\ref{Energy1}), in the geodesic integral of the motion,  (\ref{Vir3}), one gets
\begin{equation}
\bigg( \frac{d \lambda}{d \tau}\bigg)^2   ~=~ 4\, \bigg[\, (\alpha' E)^2 ~-~ \alpha'\, \langle \cN_{\rm osc} \rangle  \,  \frac{(r^2 +a^2)}{a R_y A^3 \, \Gamma^2 }  \,\bigg]\,.
  \label{lameqn1}
\end{equation}
Finally, remembering that $r$ is related to $\lambda$ via  (\ref{rlambdareln}), this gives the radial equations
\begin{equation}
\bigg( \frac{d r}{d \tau}\bigg)^2   ~=~    (\alpha' E)^2  A^2\, \Gamma^2  ~-~ \alpha'\, \langle \cN_{\rm osc} \rangle  \,  \frac{(r^2 +a^2)}{\sqrt{Q_1 Q_5} }  \,, 
  \label{radeqn1}
\end{equation}
where we have used $\sqrt{Q_1 Q_5} = R_y a A$.
For small $a \ll r, b$  this gives 
\begin{equation}
\bigg( \frac{d r}{d \tau}\bigg)^2   ~=~    (\alpha' E)^2 \, \bigg( 1 +\frac{b^2\,n}{2\, r^2} \bigg)~-~ \alpha'\, \langle \cN_{\rm osc} \rangle  \,  \frac{r^2 }{\sqrt{Q_1 Q_5} }  \,, 
  \label{radeqn2}
\end{equation}
which leads to the following estimate of the maximum return height:
\begin{equation}
r_{\rm max}   ~\approx~   \bigg( \frac{(\alpha' E)^2 \,b^2\,n\, \sqrt{Q_1 Q_5}   }{2\,\alpha'\, \langle \cN_{\rm osc} \rangle}\bigg)^{\frac{1}{4}}  ~\approx~   \bigg( \frac{(\alpha' E)^2 \,b^3 \,n\, R_y  }{2\,\sqrt{2}\, \alpha'\, \langle \cN_{\rm osc} \rangle}\bigg)^{\frac{1}{4}}  \,, 
  \label{rmax1}
\end{equation}

From the estimate~\eqref{mass estimate} of the string mass after passiing through the cap
\begin{equation}
\big\langle \cN_{\rm osc} \big\rangle = \int {dk}\, k|\beta_k|^2 \sim \textit{const.}\Big(\frac{b\,(\alpha' E)}{n\,a^2}\Big)^2 
\label{Number1}
\end{equation}
This gives:
\begin{equation}
r_{\rm max}   ~\sim~   a\, n^{\frac{3}{4}}\,  \bigg( \frac{Q_1 Q_5}{\alpha'^2}\bigg)^{\frac{1}{8}}    ~\sim~  a^{\frac{3}{4}}\, b^{\frac{1}{4}}\, n^{\frac{3}{4}}\, \bigg( \frac{a \, R_y}{\alpha'}\bigg)^{\frac{1}{4}} ~\sim~  a^{\frac{3}{4}}\, b^{\frac{1}{4}}\, n^{\frac{3}{4}} \, \sqrt{g_6} \, \big(j_L   \big)^{\frac{1}{8}}    ~.
  \label{rmax2}
\end{equation}
%
For deep throats ($j_L$ small but large enough that the classical approximation still holds)  
\begin{equation}
r_{\rm max}   ~\sim~   a^{\frac{3}{4}}\, b^{\frac{1}{4}}\, n^{\frac{3}{4}} \, \sqrt{g_6}  \,.
  \label{rmax3}
\end{equation}

Note that one can always arrange to have $g_6<1$.  To see this, observe that the duality transformation $ST_4S$, where $S$ is type IIB S-duality and $T_4$ is T-duality on all four cycles of $\IT^4$, sends $g_6\to 1/g_6$ while preserving the D1 and D5 charges.   This duality transformation is therefore a symmetry of the background and it allows us to restrict  to $g_6<1$.%
\footnote{More precisely, $g_6$ combines with the RR axion as well as the four-form with legs on $\IT^4$ to form a complex modulus that undergoes fractional linear transformations, whose fundamental domain is restricted to the weak coupling domain of $g_6$; see~\cite{Larsen:1999uk} for details.}

The first equality of~\eqref{rmax2} tells us that $r_{\rm max}\gg a$, since $\sqrt{Q_1Q_5}/\alpha'$ is the ratio of the $AdS_3$ radius to the string scale which we take to be large in the supergravity approximation; on the other hand, the last equality of~\eqref{rmax2} tells us that $r_{\rm max}\ll b$.  Thus, while the string rebounds to a height outside the cap, it is trapped far down the BTZ-like throat of the microstate geometry.

It is worth noting that for a massless particle one has $\cN_{\rm osc}=0$ and  (\ref{radeqn1}) gives $|\frac{d r}{d \tau}|    ~=~  | (\alpha' E)   A \, \Gamma|$.  The excited string is therefore highly relativistic so long as one has:
\begin{equation}
(\alpha' E)^2  A^2\, \Gamma^2  ~\gg~ \alpha'\, \langle \cN_{\rm osc} \rangle  \,  \frac{(r^2 +a^2)}{\sqrt{Q_1 Q_5} } \,.
  \label{relativisticbd1}
\end{equation}
Using  (\ref{Number1}), and assuming $a \ll b$,  this is equivalent to
\begin{equation}
  \frac{ 2\, \alpha' }{\sqrt{Q_1 Q_5}}   \, \frac{(r^2 +a^2)}{n^2\, a^2\,\Gamma^2  } ~\ll~ 1 \,.
  \label{relativisticbd2}
\end{equation}
In the cap  region, $0 \le r \lesssim \sqrt{n} a$,  one has $\Gamma \sim 1$, and this inequality is guaranteed because in the supergravity background one must have  $\sqrt{Q_1 Q_5} /\alpha' \gg 1$.  Thus the string is still highly relativistic as it crosses and leaves the cap.

It is also significant to note that at  $r \sim r_{\rm max}$, the tidal forces are of order:
\begin{equation}
\frac{a^2 \, b^2 \,  n\,  (\alpha' E)^2}{r^6} ~\sim~\bigg( \frac{b}{a^5 \, n^7 }  \bigg)^{\frac{1}{2}} \, \frac{(\alpha' E)^2}{g_6^{\,3}}  \,,
  \label{tidesrmax}
\end{equation}
which is still ``large.''  That is, our assumption that we can approximate the return trip up the throat again using the Penrose approximation is not strictly valid because the string has slowed to a halt in the region in which the tidal forces are still strong.  But qualitatively one expects a similar evolution, with large tidal forces causing mode creation that further slows the string and lowers its return height.  However, in computing the return height, we have only used the excitations generated in the trip down the throat and so any further excitations generated by tidal forces on the return journey will simply decrease the  estimated return height.  This is consistent with philosophy inherent in our other assumptions and simplifications: The estimates  (\ref{rmax2}), or  (\ref{rmax3}), represent an upper bound.

\subsection{Typical shape of the string}
\label{ss:shape}

The distribution of string mode excitations is relatively flat out to the level $k_{\rm max}$.  As the string yo-yos up and down the throat, each return pass is a bit lower as more center-of-mass kinetic energy is diverted into oscillator excitations.  The lower value of $E$ leads to a smaller ceiling $k_{\rm max}$ on the number of oscillators that are further tidally excited.  In the end, the lower modes will be excited more, as they experience more passes during which they gain additional excitation.  During the intervals that a given mode is unstable, its quantum state undergoes a squeezing transformation~\cite{Albrecht:1992kf}.  After the string has settled into the cap, it will be in some highly squeezed state of its oscillator modes up to level $k_{\rm max}$.  The string will be wiggling back and forth wildly, mostly along the diagonal in the $y$-$\varphi_2$ plane.  The typical string profile is sketched in Figure 3.

\vspace{0.5mm}
\begin{figure}[h!]
\centering
\includegraphics[width=.5
\textwidth]{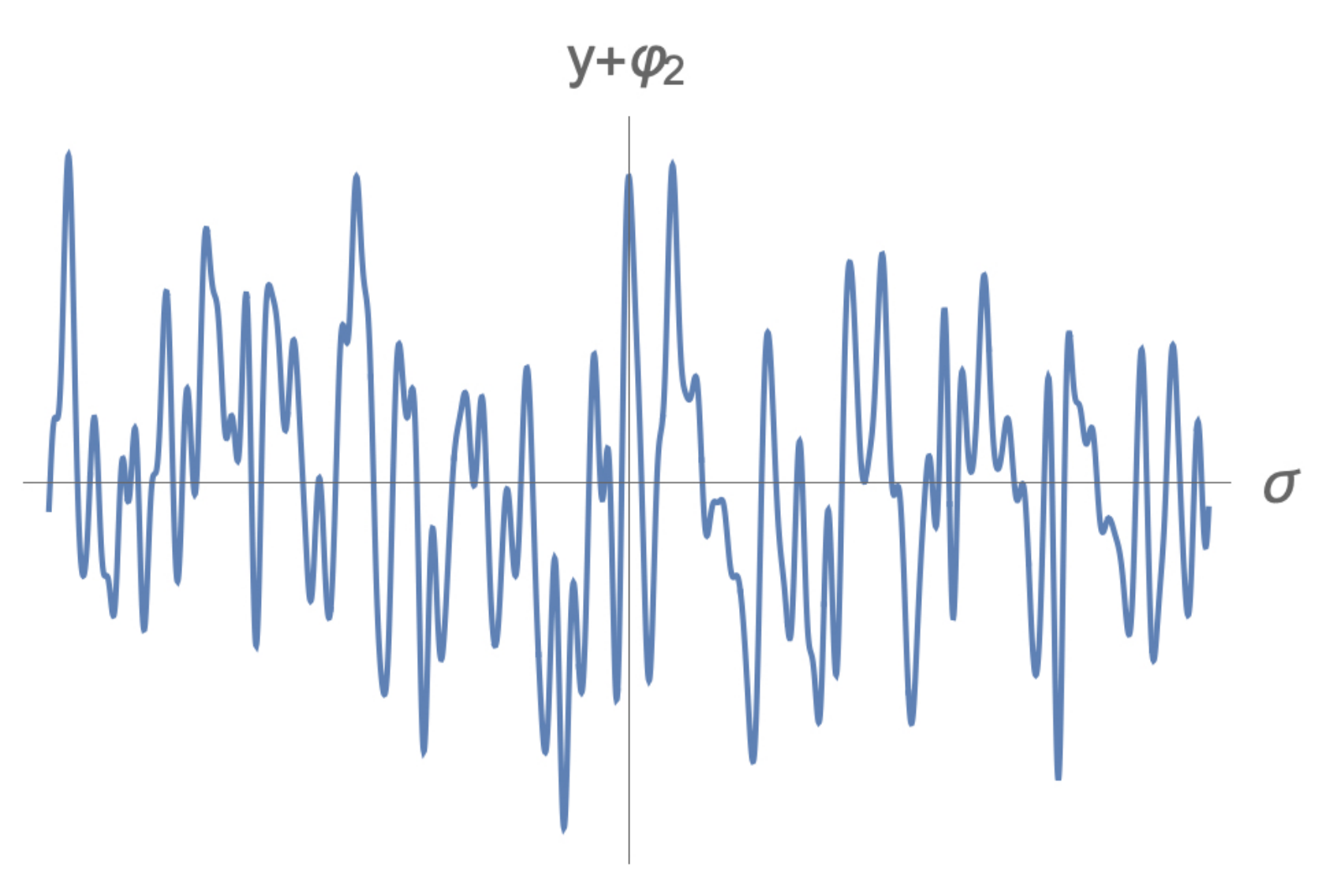}
\caption{\it Typical profile of string oscillations after tidal excitation.}
\label{fig:wiggly}
\end{figure}
\vspace{1mm}

\section{Tidal forces revisited}
\label{sec:Tides}

When  the tidal forces greatly exceed the string tension, then the string can be approximated by a line of independently-moving, free particles.  This means that  the classical stretching of the string can be modeled by integrating the equations of geodesic deviation.   Our goal here is to revisit the earlier work on tidal forces \cite{Tyukov:2017uig,Bena:2018mpb}  and do precisely this calculation.  As one would expect,  the results are in  accord with the more precise (quantum) string computations  of Section \ref{sec:stringprop}.

The equation of geodesic deviation is 
\begin{equation}
A^\mu ~\equiv~ \frac{D^2 S^\mu}{d \tau^2}  ~=~ -{\cA^\mu}_\rho \,  S^\rho  \,, \qquad {\cA^\mu}_\rho ~\equiv~  {R^\mu}_{\nu \rho \sigma} \, V^\nu \, V^\sigma \,,
  \label{Geodev}
\end{equation}
where  $V^\mu = \frac{dx^\mu}{d \tau}$ is the proper velocity of the probe and  $S^\rho$ is the deviation vector.  The matrix ${\cA^\mu}_\rho$ is sometimes known as the  ``tidal tensor''.\footnote{Note that we have reversed the sign of ${\cA^\mu}_\rho$  relative to the definition used in  \cite{Tyukov:2017uig,Bena:2018mpb,Bena:2020iyw}.  This brings the definition into line with more standard usage.}

The vector, $S^\rho$, is a space-like vector in the rest-frame of the geodesic observer and $A^\mu$ represents the acceleration per unit distance, or the tidal stress.  The skew-symmetry of the Riemann tensor means that $ A^\mu V_\mu  =0$ and so the tidal acceleration is similarly space-like, representing the tidal stress in the rest-frame of the infalling observer with velocity $V^\mu$.  Observe that negative eigenvalues of ${\cA^\mu}_\rho$ correspond to increasing separation of the geodesics, or stretching of an extended object.

We consider this for the radial, time-like geodesics at $\theta =0$ defined in Section \ref{ss:geodesics}. These geodesics are defined by the conserved quantities (\ref{ConsMom}) and (\ref{CentBarr1}), and the velocities in (\ref{thetavel}) and (\ref{radvelsq}).  
Note that in this section, $\tau$ will denote proper time along the geodesic and $m=1$ is some dimensionless reference mass, while in the previous section we used $\tau$ to denote a dimensionless coordinate time along the string worldsheet and $m$ was the string's mass.  As we remarked earlier, this means that the energy parameter will be $\geoE$, the dimensionless energy integral along time-like geodesics.

\subsection{Expansions of the tidal tensor}
\label{ss:tidalexp}

Before trying to extract the eigenvalues of ${\cA^\mu}_\rho$, it is useful to make a first pass by looking at the magnitude of the tidal tensor: 
\begin{equation}
|\cA| ~\equiv~  \sqrt{{\cA^\mu}_\rho\, {\cA^\rho}_\mu}  \,.
  \label{cAnorm}
\end{equation}
The dominant term for particles dropped from high up the throat involves $|\cA| \sim \geoE^2 b^2$.  Indeed, the complete dominant term, $|\widehat \cA|$, with this  behavior is given by:
\begin{equation}
|\widehat \cA|^2 ~=~  \frac{2\, \geoE^4 \,  b^4 \, r^{4(n-1)}}{(r^2 +a^2)^{2(n+2)} } \,  \bigg[  \, 5\,r^4 \, F^2 ~+~ 2n \,(7 n+2) \, a^2 \, r^2  \,F(r)  ~+~ n^2 (2n^2 -2 n+1) \, a^4\,   \bigg]  \,,
  \label{cAdom1}
\end{equation}
where $F$ is defined in (\ref{FGAdefns}).   For small $a$ this gives 
\begin{equation}
|\widehat \cA| ~\sim~ \frac{2 \,a^2 \, b^2 \,  n\sqrt{(n+1)(n+5)} \,  \geoE^2}{r^6} \,.
\label{domstress}
\end{equation}

To determine the tidal effects more precisely, it is essential to know the signs and magnitudes of the eigenvalues of  ${\cA^\mu}_\rho$.  The expansion of this matrix at large $b$ and small $a$ turns out to be a rather subtle process.  There the some very large ($\sim b^4$) and very small  ($\sim a^4$) off diagonal terms that need to be retained in order to extract the dominant eigenvalues.  After careful computation we find that to leading order there are two non-trivial, dominant eigenvalues:
\begin{equation}
\lambda_\pm ~=~ \frac{a^2 \, b^2 \, \geoE^2 }{r^6} \, n\,\Big(n+1 ~\pm~ \sqrt{(n+1)(n+9)}   \, \Big) \,,
\label{tidalevs1}
\end{equation}
with all other eigenvalues either exactly zero or sub-leading.  As confirmation of this, one should note that 
\begin{equation}
\lambda_+^2 ~+~ \lambda_-^2 ~=~ \frac{4 \,a^4 \, b^4 \,  \geoE^4}{r^{12}} \,  n^2 \, (n+1)(n+5) \,,
\label{check1}
\end{equation}
which is  consistent with (\ref{domstress}).  Moreover, these eigenvalues have the identical functional form to the eigenvalues, $\mu^\pm$ in  (\ref{evals1}) and  (\ref{mu evalues}),  for the mass matrix  in the Penrose limit.
We are thus seeing exactly the same physical stress to that experienced by the string.

In particular, $\lambda_+$ is positive and  $\lambda_-$ is negative, and  for large $n$  one has
\begin{equation}
\lambda_+ ~\sim~ \frac{2\, n(n+3) \,a^2 \, b^2 \, \geoE^2 }{r^6} \,, \qquad \lambda_- ~\sim~ - \frac{4\, n\,a^2 \, b^2 \, \geoE^2 }{r^6}  \,.
\label{tidalevs2}
\end{equation}
From the signs in (\ref{Geodev}) one sees that $\lambda_+$ generates a compression in an extended object, while  $\lambda_-$ a tension.  This is also consistent with the physics of positive and negative effective masses for the stringy modes and so we have a useful consistency check on the signs of the string mass matrix.

It is also useful to estimate the tidal force in the cap.  The peak value of (\ref{cAdom1})  arises near $r \sim \sqrt{n}\, a$  and is approximately  
\begin{equation}
|\widehat \cA|_{r  = \sqrt{n} a}  ~\sim~  \frac{2\, \geoE^2 \,  b^2 }{ a^4\, e\, n} \,  Q(n)  \,,
  \label{cApeak}
\end{equation}
where  $e \approx  2.71828 ...$ is Euler's number, and  $Q(n)$ is a positive function of $n$ that limits to  $1$ at large $n$.   

It is significant to observe that if one puts  $r = \sqrt{n}\, a$ in (\ref{domstress}), the result is the same as (\ref{cApeak}), up to terms of  order unity.  Thus the small $a$ approximation works well (at least parametrically) down to the ``edge of the cap'' at $r = \sqrt{n}\, a$.

For $n>1$,  the dominant tidal force, $|\widehat \cA|$, vanishes at $r=0$.  Indeed the complete expression for $| \cA|$ reduces to 
\begin{equation}
| \cA|_{r =0}  ~=~\frac{\sqrt{2}}{R_y \, \sqrt{a^2 + \frac{1}{2} b^2}  }   ~=~\frac{\sqrt{2}}{\sqrt{Q_1 \, Q_5}}  ~,
  \label{cAcenter}
\end{equation}
which is the exact expression for the tidal tensor while falling into the BTZ, or AdS$_3$, metric.  Thus, for $n>1$, one should think of the tidal force in the cap as peaking at the edge of the cap and then dying away to its low AdS  value in the center. For $n=1$, the tidal force stays roughly at its peak value across the entire cap.

\subsection{Integrating the geodesic deviation}
\label{ss:intgeodev}

\def\muparam{\nu}
\def\nuparam{\rho}

In much the same spirit as the computation of the excitations of the string, one can make a classical estimate of the stretching effect of the negative eigenvalue, (\ref{tidalevs1}), of the tidal tensor.  

We consider the radial, time-like geodesics described in Section \ref{ss:geodesics}.   Setting $m=1$ in   (\ref{radvelsq})  and expanding in small $a$, ($a \ll r, b$),  the radial equation  becomes 
\begin{equation}
r^2 \bigg(\frac{dr}{d\tau} \bigg)^2 ~=~ \geoE^2 (2 r^2 + n\, b^2) ~-~ \frac{\sqrt{2}}{b\,R_y} \, r^4
\label{radeqn}
\end{equation}
This is elementary to integrate and gives 
\begin{equation}
r^2  ~=~ 2 \sqrt{2} \, b\,R_y \, \geoE^2 (1 ~+~  \nuparam \, \cos \muparam \tau )\,, \qquad \muparam ~\equiv~ \frac{2^\frac{1}{4}}{\sqrt{b \,R_y}}\,, \qquad \nuparam ~\equiv~  \sqrt{1 + \frac{n\, b}{2 \sqrt{2} \, \geoE^2\,R_y}}\,.
\label{radevol}
\end{equation}
The energy, $\geoE$, sets the amplitude of the motion and the time-scale for infall is $\muparam^{-1} \sim \sqrt{b R_y}$.  On the face of  it, this is a ``long time'' measured in units of $a$ or $R_y$.

One can obtain a crude approximation to the tidal deformation of an object by solving: 
\begin{equation}
 \frac{d^2 s}{d \tau^2}  ~=~ -\lambda_- ~=~  \frac{4\, n\,a^2 \, b^2 \, \geoE^2 }{r^6}\,.
  \label{Geodevapprox1}
\end{equation}
This approximation is reasonable provided that the eigenvector does not rotate significantly during the infall.  Indeed, the string computation shows that this is true in the throat all the way to the edge of the cap. 

Using (\ref{radevol}), this equation becomes:
\begin{equation}
 \frac{d^2 s}{d \tau^2}  ~\sim~  \frac{ a^2 \, n}{b\,R_y^3 \, \geoE^4 (1 ~+~  \nuparam \, \cos \muparam \tau )^3}   \,.
  \label{Geodevapprox2}
\end{equation}
If  one releases the probe from $r \sim b$, then $\geoE^2 \sim b/R_y$ and then one has:  
\begin{equation}
r^2  ~\sim~  b^2  (1 ~+~  \nuparam \, \cos \muparam \tau )\,, \qquad  \frac{d^2 s}{d \tau^2}  ~\sim~  \frac{ a^2 \, n}{b^3 \,R_y (1 ~+~  \nuparam \, \cos \muparam \tau )^3}   \,.
\label{radevol2}
\end{equation}
``Half-way down" the throat, $r \sim \sqrt{a b}$, corresponds to
\begin{equation}
(1 ~+~  \nuparam \, \cos \muparam \tau )  ~\sim~  \frac{a}{b}  \,, \qquad \Rightarrow \qquad  \frac{d^2 s}{d \tau^2}  ~\sim~  \frac{ n}{a \,R_y}   \,. 
\label{radevol3}
\end{equation}
Using (\ref{jLbound}), we see that at this point the tidal force is reaching the Planck/compactification scale   \cite{Tyukov:2017uig,Bena:2018mpb}.  However, when measured using the proper time on the geodesic, this occurs very late in the infall.  Indeed, the time to travel between $r \sim \sqrt{a b}$ and  the bottom of the throat, $r \sim a$, is approximately 
\begin{equation}
\Delta \tau   ~\sim~  \frac{1}{\muparam \nuparam } \, \frac{a}{b} ~\sim~   \frac{a \, R_y \, \geoE}{b \sqrt{n }} ~\sim~   \frac{a}{ \sqrt{n }}  \,,
\label{Deltatau}
\end{equation}
which is an extremely short time. So while the force is huge, the time over which it acts is small. The probe does indeed encounter super-Planckian tidal forces, but this only happens once the probe is ultra-relativistic and then the time dilation means that the effects of these forces remain small.  

This suggests that one should drop the probe in the region in which the tidal forces are large so that it spends  a longer proper time encountering these forces.  However, one should remember that a significant contribution to the large tidal forces came through  relativistic magnification factor of $\geoE^2 \sim b/R_y$ in  (\ref{domstress}) and this magnification will be diminished by releasing the probe further down the throat.  We therefore investigate the effects of the tidal forces more carefully.

It is straightforward to integrate   (\ref{Geodevapprox2}) once to obtain
\begin{equation}
\begin{aligned}
 \frac{d s}{d \tau}  ~\sim~  \frac{a^2 \, n }{b\,R_y^3 \, \geoE^4\, \muparam}\, \bigg[ \frac{\nuparam^2 +2}{(\nuparam^2 -1)^\frac{5}{2}} & \, {\rm arctanh}\bigg(  \sqrt{\frac{\nuparam-1}{\nuparam+1} } \, \tan\big(\coeff{1}{2} \muparam \tau \big) \bigg) \\
&   ~+~\frac{\nuparam}{2(\nuparam^2 -1)^2} \, \frac{(1 - 4 \nuparam^2 - 3\nuparam \, \cos \muparam \tau )  }{(1 ~+~  \nuparam \, \cos \muparam \tau )^2} \,\sin \muparam \tau \bigg]  \,,
 \end {aligned}
  \label{Geodevapprox3}
\end{equation}
where we have chosen the constant of  integration so that $ \frac{d s}{d \tau}$ vanishes at  $\tau =0$, where $r$ takes its maximal value. One can also integrate this again and obtain a result expressed in terms of dilogarithms, however it is simpler to observe that the second term in   (\ref{Geodevapprox3}) gives the dominant contribution deeper in the throat, and this is trivially integrated to give 
\begin{equation}
\begin{aligned}
\Delta s  ~\sim~ &  \frac{a^2  \, n}{b\,R_y^3 \, \geoE^4\, \muparam^2}\,\frac{1}{2(\nuparam^2 -1)} \, \bigg[ \frac{1}{(1 ~+~  \nuparam \, \cos \muparam \tau )} ~+~ \frac{3}{(\nuparam^2 -1)}\, \log(1 ~+~  \nuparam \, \cos \muparam \tau ) \bigg]  \\
 ~\sim~  & \frac{a^2}{b\,R_y \, \geoE^2} \,\frac{1}{(1 ~+~  \nuparam \, \cos \muparam \tau )} ~\sim~   \frac{a^2}{r^2} \,,
\end {aligned}
  \label{Geodevapprox4}
\end{equation}
where we have used (\ref{radevol2}). 

This means that, independent of the value of $\geoE$, or release point of the probe, the tidal displacement per unit length, becomes of order $1$ (or, more precisely, $\frac{1}{n}$) only when the probe reaches the edge of the cap,  $r \sim \sqrt{n} a$. 

It  is also very instructive to estimate the tidal distortion delivered by passage though the cap.  The tidal tensor has a peak value modulus given by (\ref{cApeak}):  
\begin{equation}
 \frac{\geoE^2\,  b^2}{n\, a^4}\,.
\label{tidal-cap}
\end{equation}
While the eigenvalue analysis in the cap is extremely complicated, we suspect that the negative eigenvalue will, at worst, get a factor of $\frac{1}{n}$.    
For an infalling geodesic, the dominant term in the radial velocity is
\begin{equation}
\frac{d r}{d \tau} ~\sim~  \frac{b\, \geoE}{a}  \,,
\label{radvel-cap}
\end{equation}
and hence, in the crossing the cap one has:
\begin{equation}
 \frac{d^2 s}{d r^2}  ~\sim~ \frac{1}{n}  \times  \frac{1}{n \,a^2}   \qquad \Rightarrow \qquad \Delta s ~\sim~ \frac{(\Delta r)^2}{n^2\,a^2} ~\sim~\cO(n^{-1}) \,,
  \label{distortion2}
\end{equation}
for $\Delta r \approx \sqrt{n} \,a$.

Thus, in crossing the cap, the forces are, in principle, super-Planckian but in reality they deliver an impulse that creates a displacement of $\cO(n^{-1})$.  Also note that the factors of $\geoE$ have cancelled and so the result is independent of where the probe is released.

The end result of integrating the geodesic deviation is thus consistent with the string computation.  The tidal forces go well above the string tension and so overwhelm the tension and stretch the string.  For the lowest string modes, the string may be thought of as a line of particles and its distortion can be approximated by geodesic deviation.  The result is a a stretching of the string by an amount of order $\frac{1}{n}$.  This is consistent with the much more precise stringy result in Section \ref{ss:stringexcitation}.

\section{Final comments}
\label{sec:Conclusions}

The  fact that microstate geometries cap off at very high red-shifts leads to  interesting new physics.  In  particular, the presence of the cap leads to non-trivial multipole moments in the tidal tensors for infalling particles.  These tidal terms can reach the string/compactification scale  a significant distance above the  cap, especially when these multipole moments are amplified by the ultra-relativistic speeds that will typically be attained by a probe  \cite{Tyukov:2017uig,Bena:2018mpb,Bena:2020iyw}.   One might therefore expect that a probe  would be completely disrupted.  However we have shown that, on a single pass through  the cap of the geometry,  the large tidal forces only act over a very short period of time and this produces a limited excitation of the string modes.  Indeed, the level of string mode that becomes  excited depends on the infalling energy of the probe, and the occupation number of each excited mode is $\cO(1)$.  We show that the fleeting nature of the interaction with the tidal forces leads to a controllable approximation in which one can estimate the outgoing state of the string and  treat the excited state as a massive object.

The end result of the tidal interaction is that a significant fraction of the kinetic energy of the string probe is transferred into string excitations, giving the string a non-trivial rest mass.  This means that, with each pass through the geometry, a stringy probe becomes ever more deeply trapped in the microstate geometry.  We computed an upper bound,  (\ref{rmax intro}), on the return height after a first pass and find that it is far short of the original release height.    One of the exciting  features of this process is that it provides a mechanism through which microstate geometries generically capture string probes without having to consider the back-reaction of the probe on the geometry. The computation of such back-reaction is, of course, important, but such calculations are very challenging. The beauty of the tidal mechanism is that it is far more easily computable within a controlled approximation.   

Our results also  have the important consequence that stringy probes of deep microstate geometries will not produce sharp echoes in the linearized approximation.  Superstrata have proven extremely useful in that one can compute Green functions in these backgrounds  \cite{Raju:2018xue,Bena:2018bbd,Bena:2019azk}. These geometries have been extensively probed using scalars and one finds that signals ultimately rebound off the cap. However, in string theory such scalar fields are simply effective descriptions of strings and so our results show that there will, in fact, be no such echoes at leading order.  At non-linear order, the probe will radiate as it oscillates up and down while it settles into the cap.  One effect will be ``gravitational brehmsstrahlung''~-- as the string decelerates due to oscillator mode production, it will shed as classical radiation the gravitational field it carries along with it.  It would be interesting to determine the radiation pattern from the probe during this initial relaxation phase.

The excitation of string modes means that the  probe must fall back, and undergo yet more tidal excitation with each subsequent pass through the cap.   There are a number of relaxation modes of the string, as well as the background, once the probe has settled into the cap.  The probe will thermalize, both by emission of short strings as well as by higher-order interactions among the modes on the string (these interactions are truncated away in the Penrose approximation).  Among the available decay channels are soft BPS and near-BPS supergravity modes of the background that modify the superstratum background.  These can alter both the BPS gravitational wave profile on the superstratum, as well as the shape profile of the underlying supertube upon which the superstratum is built by adding such a supergravity wave.  One of the possible supertube shape profile deformations is to offload its angular momentum $j_{L,R}$ into angular excitations further up the throat, sending the parameter $a\to 0$; then the cap recedes to infinite redshift and the supergravity approximation loses control of the horizon-scale structure.  Another possibility is that shape gyrations of the underlying onebrane-fivebrane supertube cause it to self-intersect, at which point D-brane excitations -- the manifestation of little strings in this duality frame -- become light and again cause the breakdown of both the supergravity approximation as well as string perturbation theory~\cite{Martinec:2020gkv}.

It appears that charged/rotating black holes have a strong gravitational shock wave at their inner horizon~\cite{Marolf:2011dj,Eilon:2016osg,Chesler:2018hgn,Burko:2019fgt,Dias:2019ery,Pandya:2020ejc,Emparan:2020rnp}.  Probes reaching this structure will be tidally torn apart, absorbed and scrambled into the black hole microstate structure.  
For extremal black holes, the inner and outer horizons coincide from the perspective of the probes we are considering.  One might regard microstate geometries as a slight regularization of the near-horizon structure, replacing it by a timelike cap.  The arbitrarily large tidal forces at the inner horizon are supplanted by large but finite tidal forces at the cap; string theory probes are disrupted by these tidal forces, trapped and scrambled.  It would be interesting to understand how closely this process models absorption into a generic extremal black hole microstate.  But qualitatively there is some resemblance to the picture of absorption in the black-hole degrees of freedom of the spacetime CFT, in which a supergraviton probe is absorbed into counter-propagating momentum waves~\cite{Callan:1996dv,Das:1996wn}.
The infalling probe settling into the cap of a microstate geometry is stretched into a large string stretched along the $AdS_3$ and $\IS^3$ angular directions (see Figure~\ref{fig:wiggly}); once the string stops stretching and starts oscillating again, it will oscillate back and forth in these directions,   The radiation from such a string will be predominantly beamed in the direction of the stretching~\cite{Damour:2001bk}.  Such a tidal stretching, disruption and eventual absorption into the black hole ensemble would be the fate of a string probe in a generic black hole microstate; probe absorption by a BPS microstate geometry provides an explicit, calculable model of this process.

There are some features missing from the perturbative string dynamics in geometrical microstates, that one expects to be important in generic black hole microstates.  The generic excitations of the cap are stringy, as one sees, for example, in the enumeration of perturbative string vertex operators (see~\cite{Martinec:2018nco,Martinec:2020gkv} for a recent discussion in the context of microstate geometries); furthermore, even such perturbative string excitations seem to miss out on the additional fractionation of quanta by the underlying branes in the background~-- branes in isolation outside the horizon appear to fractionate into smaller constituents upon being absorbed by the black hole, and the generic such fractionation seems difficult to accommodate within the context of geometrical backgrounds~\cite{Shigemori:2019orj}.  These properties of generic black holes are reflective of the strongly coupled little string theory dynamics that accounts for the black hole entropy%
~\cite{Strominger:1996sh,Maldacena:1996ya,Martinec:2019wzw} of the onebrane-fivebrane system.  The bottom line is that it will be difficult to explore the absorption and scrambling processes of generic black hole microstates if the generic microstate structure involves additional stringy ingredients; however we regard it as intriguing that these finely tuned coherent microstates exhibit behaviors that closely model what is expected when a probe is absorbed into a black hole.  

Our results here suggest several interesting future projects.  On the technical level, one can try to improve upon that approximations we have made to arrive at our broad result.  Microstate geometries have many control parameters that can limit the interactions and perhaps improve precision.  For example, we have been largely motivated by the study of extremely deep throats with $b \gg a$ and $j_L$ small. It would be interesting to consider large, or intermediate values of $j_L$, where the probes are not so highly relativistic. We have also not considered the dependence on the parameter $n$ in any detail.  It is evident from (\ref{mass estimate}) that the extent of the string excitation decreases at large $n$.  This could lead to a more adiabatic approximation to  the excitation process. 

Then there are other trajectories and other microstate geometries.  We chose a radial trajectory with $\theta =0$ as a matter of computational convenience; other radial trajectories may well have more extreme tidal interactions as they enter the cap.  There are other interesting microstate geometries with richer microstate structure that could lead to even stronger tidal disruptions \cite{Mayerson:2020tcl}.  At the other extreme, it is also possible to ``soften'' the tidal multipoles \cite{Bena:2020iyw} and this may produce a weaker tidal excitation of string probes. By delaying the onset of string tidal forces one might further limit the impulse generated by strong tides, or perhaps confine the regime of strong tides to a small neighborhood of the cap.  

More broadly, it would be interesting to analyze the complete superstring, including world-sheet fermions, along with the interactions with the background tensor gauge fields.  Here we only considered the $B$ field, and showed that its influence is sub-leading.  The tensor gauge fields could play a larger role in other backgrounds. One would expect that the ``Lorentz force'' contributions of the $B$ field to generate a form of tumbling motion of an excited string, and  this would extract yet more kinetic energy from the center of mass motion.  It therefore seems very likely that our estimates of the return height are very much an upper bound because there are so many other channels in which an infalling string can become excited.

It would also be interesting to study orbits of strings in the cap, especially those near the evanescent ergosphere.  Such orbits have played a significant role in suggesting that there are potential gravitational instabilities in microstate geometries \cite{Eperon:2016cdd}.  However, it seems more likely that these apparent instabilities will actually evolve into the stringy scrambling of probes  \cite{Marolf:2016nwu,Bena:2018mpb,Bena:2020yii,Martinec:2020gkv}. It would be useful to use the methods we describe here to obtain an estimate of the rate of such stringy scrambling.  

There are also many other possible brane probes.  One would expect D1 and D5 brane probes to have a very soft  interaction with superstrata because such probes are mutually BPS with respect to the branes dissolved in the background.  One could softly break the supersymmetry to create a near-BPS probe, perhaps by tilting it, or giving it an additional, very small non-BPS charge.   

It is also tempting to make conjectures about the universality, and perhaps even observable consequences, of the results presented here.  In this paper, the tidal forces that cause the string disruption are the direct result of the non-trivial cap and a large red-shift between the cap and the top of the black-hole throat. More broadly, any deviation from the horizon geometry of a black hole should produce a similar effect, particularly if it involves an extreme red-shift between the new horizon-scale geometry and infinity.  We expect that the effect we have analyzed here for microstate geometries to be a feature of any gravitational description of a coherent expression of microstate structure. (Such a conclusion is also strongly supported by the results \cite{Gibbons:2013tqa}.)  It therefore seems very likely that horizon scale microstructure will inevitably lead to string scrambling of infalling matter at the horizon scale; even for extremal (zero-temperature) black holes.  

Naively, this would suggest that there will be no return echoes from any reasonable stringy description of horizon-scale black-hole microstructure.  However, the  tidal stress produces a highly excited string and, as we pointed out, the string  will fragment, preferentially, into light closed-string excitations.  We therefore expect such ``gravitational brehmsstrahlung'' to be emitted by the excited probe as it approaches and enters the cap, and as it slows down.  Moreover, since the probe is still traveling at relativistic speeds during much of this process, a significant fraction of the gravitational brehmsstrahlung will be beamed along the string  trajectory.  Subsequent passes through the cap will produce similar, but much diminished and more diffuse, bursts of radiation.   We therefore expect a series of progressively more muffled echoes  from the excited probe as it slows down and settles into the cap.%
\footnote{
There is already an extensive literature on the possibility of gravitational wave echoes (see for instance~\cite{Cardoso:2019rvt,Abedi:2020sgg} for reviews).  These investigations tend to focus on phenomenological issues, whereas our interest here is in a particular top-down model whose black-hole microstructure might exhibit such a phenomenon.  Our work is based on microstate geometries that are string-theory backgrounds that can support a wide range of microstate structure (though admittedly current constructions involve backgrounds far from the one we inhabit).  However, in the arena of phenomenology, we note that the fact that the probe string becomes highly excited means that the echoes, if any survive, would not simply be made up of gravitons but could involve any low-mass string excitation, and thus consist of a stream of any (and all) stable, low-mass particles in the appropriate extension of the standard model.  
}

Note that for this radiation to escape the throat, the energy of its constituent strings must be low enough that they don't themselves get tidally disrupted as they try to escape the throat; if they do, they will simply become additional massive strings that get stuck in the throat and fall back into the cap.  An estimate of the upper bound on this energy per particle comes from putting $k_{\rm max} =1$ in~\eqref{kmax} (perhaps with some safety margin to ensure that the likelihood of tidal disruption is small)
\be
E_{\rm max} ~\sim~  \frac{a^2n}{b\,\alpha'} ~\sim~  \frac{ g_6^{\,2}}{R_y} \, \frac{n \, j_L}{\sqrt{ n_1n_5} }
\label{Ecut-off}
\ee
where we have used~\eqref{eq:quantizedcharges}.    Thus any return echo will have an energy cut-off determined by (\ref{Ecut-off}). We note that the  quantity $E_{\rm max} R_y$ is  the $L_0$ eigenvalue in the dual CFT, and we recall that the lowest energy ``gap'' in the CFT is approximately $ \frac{ j_L}{n_1n_5}$, and so, while small, this cut-off energy is much higher than the energy gap. 

A distant observer would see the  gravitational brehmsstrahlung as a very faint shower of stable, light particles. The universal feature will be graviton emission,  and it would be interesting to compare the features of the first burst of gravitons  with the expected emission of gravitational waves as a probe drops into a classical black hole in general relativity.  The particle shower will also, in principle, include some of the ``usual suspects'' from the Standard Model, as well as stable, low-mass particles, like some of  the Dark Matter candidates,  that go beyond the Standard Model. The details will depend very much on the low-energy phenomenology of the underlying string theory and we will not speculate further.  We will however note that the microstate geometry, like a black hole, is indeed acting like a particle accelerator, and any observable details of gravitational brehmsstrahlung will be sensitive to BSM physics, but unlike a black hole, a microstate geometry affords the possibility of measuring the output of the accelerator.

Of course all of these potentially observable effects must be taken with a grain of salt, given the non-generic structure of microstate geometries. It may well be that the dominant decay channels for probes of more generic microstates are rather different. However, the hope is that microstate geometries are providing a glimpse of the stringy effects underlying the dissipative processes involved in the scrambling of matter into black holes.

\section*{Acknowledgments}
\vspace{-2mm}
EJM thanks IPhT Saclay for hospitality during the course of this work; in addition, we both thank the Centro de Ciencias de Benasque for hospitality during the 2019 workshop {\it ``Gravity - New perspectives from strings and higher dimensions"}, and for the opportunity to present our work.  NW would also like to thank the organizers of the following conferences for the opportunity to present our work: {\it Holography, Gauge Theories, and Black Holes"},   Southampton,  June 2019, {\it ``Supergravity, 2019''} Padova, September, 2019;  {\it ``Geometry from the Quantum''},  KITP Santa Barbara,  January, 2020.  The work of EJM is supported in part by DOE grant DE-SC0009924. The work of NW was supported in part by the ERC Grant 787320 - QBH Structure, and by the DOE grant DE-SC0011687. 

While we were finishing the write-up of our results, we learned of work of M. Dodelson and H. Ooguri~\cite{Dodelson:2020lal} employing similar methods to study rather different physics, namely the singularity structure of the boundary two-point correlator in AdS black hole backgrounds.  While their interest is in the likelihood that the string remains unexcited along trajectories that remain outside the black-hole photon sphere, our focus has been on the highly-excited wave-functions of strings traveling trajectories that dive deeply into microstate geometries. 

\vspace{2cm}
%

\begin{appendix}

\section{The {\textit B}-field flux} 
\label{app:Bfield}

The superstratum is supported by magnetic fluxes in the three underlying tensor gauge fields in six-dimensions.  Two of these fluxes are associated with the D1 and D5 brane sources, while the third, anti-self-dual, flux  is associated with the F1-NS5 sources.  Specifically, this flux in six-dimensions comes from the trivial compactification of the Kalb-Ramond field, $B_{\mu \nu}$, of the NS sector of the IIB theory in ten dimensions.  This field is therefore essential to our analysis because it couples directly to the string probe.  These fluxes are obtained directly by solving the BPS equations, and may be found in several places \cite{Shigemori:2013lta,Bena:2015bea,Bena:2016ypk,Bena:2017xbt,Heidmann:2019zws}. We will catalogue the essential details here.

\subsection{Building the fluxes}
\label{app:build}

One writes the six-dimensional metric in its canonical BPS form \cite{Gutowski:2003rg}:
\begin{equation}
ds_6^2 ~=~    -\frac{2}{\sqrt{\cP}} \, (dv+\beta) \big(du +  \omega + \tfrac{1}{2}\, \mathcal{F} \, (dv+\beta)\big) 
~+~  \sqrt{\cP} \, ds_4^2(\cB)  ~\equiv~ g_{MN} dz^M d z^N \,. \label{sixmetGMR}
\end{equation}
The metric, $ds_4^2$, on the four-dimensional base, $\cB$, is that of flat $\IR^4$, written it in the standard bipolar form:
\begin{equation}
ds_4^2 ~=~ \Sigma\, \bigg( \frac{dr^2}{(r^2 + a^2)} ~+~ d \theta^2 \bigg)  ~+~ (r^2 + a^2) \sin^2 \theta \, d\varphi_1^2 ~+~ r^2  \cos^2 \theta \, d\varphi_2^2   \,,
\label{basemet}
\end{equation} 
where 
\begin{equation}
\Sigma ~\equiv~ (r^2 + a^2 \cos^2 \theta) \,.  
\label{Sigdefn}
\end{equation}
We also define 
 \begin{equation}
\beta ~=~  \frac{R_y \,a^2}{\sqrt{2}\,\Sigma}\,(\sin^2\theta\, d\varphi_1 - \cos^2\theta\,d\varphi_2)\,, \qquad \Theta^{(3)}  ~\equiv~ d\beta  \,.  \label{betadefn}
\end{equation}
The coordinates $(u,v,r,\theta, \varphi_1, \varphi_2)$ are precisely those of Section \ref{sec:MG}, and the metric (\ref{sixmetGMR}) is the same as  (\ref{sixmet1})  provided that one takes 
 \begin{equation}
 \begin{aligned}
\cP ~=~  &  \frac{Q_1 Q_5}{\Sigma^2} \, \Lambda \,,\qquad \cF ~=~   - \frac{b^2}{a^2}\, F \,, \\
\omega ~=~ &  \frac{R_y\, a^2}{\sqrt{2} \, \Sigma} \, ( \sin^2 \theta \, d\varphi_1 +   \cos^2 \theta \, d \varphi_2 )  ~+~   \frac{R_y\, b^2}{\sqrt{2}\,  \Sigma}\, 
F \, \sin^2\theta\, d\varphi_1\,, 
 \end{aligned}
  \label{PFomres}
\end{equation}
with $F$ and $\Lambda$ defined in (\ref{GamLamdefns}).

To define the fluxes, it is convenient to introduce the frames:
\begin{equation}
e_1 ~=~ \frac{\Sigma^{1/2} }{(r^2 + a^2)^{1/2}} \, dr\,, \quad   e_2 ~=~\Sigma^{1/2}   \, d\theta\,, \quad   e_3 ~=~(r^2 + a^2)^{1/2}  \sin \theta  \, d\varphi_1\,, \quad   e_4 ~=~ r  \cos  \theta \, d\varphi_2 \,,
\label{frames1}
\end{equation} 
and the self-dual two-forms on $\cB$:
\begin{equation}\label{selfdualbasis}
\begin{aligned}
\Omega^{(1)} &\equiv \frac{dr\wedge d\theta}{(r^2+a^2)\cos\theta} + \frac{r\sin\theta}{\Sigma} d\varphi_1\wedge d\varphi_2 ~=~  \frac{1}{\Sigma \, (r^2+a^2)^\frac{1}{2}  \cos\theta} \,(e_1 \wedge e_2 +  e_3 \wedge e_4)\,,\\
\Omega^{(2)} &\equiv  \frac{r}{r^2+a^2} dr\wedge d\varphi_2 + \tan\theta\, d\theta\wedge d\varphi_1  ~=~  \frac{1}{\Sigma^\frac{1}{2}\, (r^2+a^2)^\frac{1}{2} \cos\theta} \,(e_1 \wedge e_4 +  e_2 \wedge e_3)   \,,\\
 \Omega^{(3)} &\equiv \frac{dr\wedge d\varphi_1}{r} - \cot\theta\, d\theta\wedge d\varphi_2~=~  \frac{1}{\Sigma^\frac{1}{2}\, r \sin\theta} \,(e_1 \wedge e_3 -  e_2 \wedge e_4)  \,.
\end{aligned}
\end{equation}
One then defines a potential, $Z_4$, and a magnetic flux,  $\Theta_4$, by
\begin{equation}
\begin{aligned}
Z_4  ~=~ &   \frac{b\,   R_y}{\Sigma} \;\! \Delta_{1,0,n}   \;\! \cos \chi_{1,0,n} \,,  \\
\Theta^{(4)}   ~=~ &  {\sqrt{2}} \,b \,n \,  \Delta_{1,0,n} \, \bigg[\Big(  \frac{\Sigma}{r \, \sin \theta} -r\, \sin \theta  \Big)\, \sin \chi_{1,0,n} \,\Omega^{(1)}  ~+~ \cos  \chi_{1,0,n} \, \Omega^{(3)} \bigg] \,,
\end{aligned}
\label{Z4Theta4form}
\end{equation}
where
\begin{equation}
\label{SSmodes1}
\chi_{k,m,n} ~\equiv~  \coeff{\sqrt{2}}{ R_y}\, (m+n) \, v ~+~  (k-m) \, \varphi_1 ~-~   m \,  \varphi_2    \,,
\end{equation}
and
\begin{equation}\label{Deltadefn}
\Delta_{k,m,n}~\equiv~ \frac{a^k \, r^n }{(r^2+a^2)^{\frac{k+n}{2} }}\,\sin^{k-m}\theta\,\cos^m\theta\,.
\end{equation}

From these building blocks one then assembles the NS two-form potential \cite{Giusto:2013rxa,Bena:2015bea}: 
\begin{equation}
B ~=~ -\frac{Z_4}{\cP}\,(d u+\omega) \wedge(d v+\beta) ~+~  a_4 \wedge  (d v+\beta) ~+~  \delta_2\,, 
\label{Bform1}
\end{equation}
where $a_4$ and  $\delta_2$ are defined by:
\begin{equation}
\label{Thetarels}
\Theta_4 \equiv \mathcal{D} a_4 ~+~ \dot{\delta_2}\,.
\end{equation}
The lack of uniqueness in determining $a_4$ and $\delta_2$ reflects the gauge freedom $B  \to B + d \Lambda$.    Since the flux depends on $v$ through a phase, the simplest choice is to take $a_4 =0$ and  integrate $\Theta^{(4)}$ to obtain $\delta_2$:
\begin{equation}
\delta_2 ~=~  b \,R_y \,  \Delta_{1,0,n} \, \bigg[ \Big( r\, \sin \theta - \frac{\Sigma}{r \, \sin \theta}  \Big)\, \cos \chi_{1,0,n} \,\Omega^{(1)}  ~+~ \sin  \chi_{1,0,n} \, \Omega^{(3)} \bigg]  \,.
\label{deltaform1}
\end{equation}
Having made these choices, we also make a gauge transformation and  take:
\begin{equation}
B ~=~ -\frac{Z_4}{\cP}\,(d u+\omega) \wedge(d v+\beta) ~+~  \delta_2~+~  d_6 \bigg[ \,\frac{r_y \, b}{r}\, \Delta_{1,0,n} \, \big( \cos \chi_{1,0,n} \,   dr ~-~ r\, \sin \chi_{1,0,n} \, d \varphi_2 \big)\bigg]\,, \label{Bform2}
\end{equation}
where $d_6$ is the full exterior derivative on the six-dimensional manifold.  This gauge choice has the virtue of having a smooth Penrose-G\"uven limit.

\subsection{The Penrose-G\"uven limit of the {\textit B}-field}
\label{app:PenroseB}

Again we will follow \cite{Blau:2002mw} in taking the Penrose-G\"uven limit:
\begin{equation}
\begin{aligned}
B_{MN} ~\to~ \Omega^{-2}\,B_{MN} \,; \qquad   & \lambda  ~\to~  \lambda  \,, \quad   t  ~\to~ t_0 + \Omega^{2}\, t  \,, \quad y~\to~ y_0 + \Omega \, y \,, \\
 &  \varphi_1~\to~ \varphi_1\,, \quad   (\theta, \varphi_2)  ~\to~ \Omega\, (\theta, \varphi_2)  \,,
\end{aligned}
  \label{PenscalingB}
\end{equation}
where we have introduced initial positions, $(t_0, y_0)$ because the fluxes depend upon $t$ and $y$.  This limit can lead to divergences that are pure gauge and this is why we made the gauge transformation in (\ref{Bform2}).

Applying the coordinate transformation to $x^{\mu'} \equiv (\lambda, t,y, \theta, \varphi_1, \varphi_2)$ as in (\ref{nullvars1}), and taking the limit   (\ref{PenscalingB}), we obtain 
\begin{equation}
\begin{aligned}
B ~\to~   \frac{ b\, A \, \theta \, \Gamma\, r^{n-1}}{2\, (r^2 +a^2)^{\frac{1}{2} (n+1)}} \,  &\Big( r\, \Gamma \, \cos  \chi^{(0)}_{1,0,n} + a\, n \, \sin  \chi^{(0)}_{1,0,n}  \Big)  \, d\lambda \wedge dy  \\ 
~-~ \frac{ a\,b\, A \, R_y\, \theta \,  r^{n-1}}{2\, (r^2 +a^2)^{\frac{1}{2} (n+3)}} \, &\bigg[ a\, \Gamma\,\Big( r\, \Gamma \, \cos  \chi^{(0)}_{1,0,n} + a\, n \, \sin  \chi^{(0)}_{1,0,n}  \Big)  \\ 
& ~+~  r\, \Big(-r\, \Gamma \, \sin  \chi^{(0)}_{1,0,n} + a\, n \, cos  \chi^{(0)}_{1,0,n}  \Big) \bigg]\,   d\lambda \wedge d\varphi_2\,, 
\end{aligned}
\label{Blim1}
\end{equation}
where $\chi^{(0)}_{1,0,n}  = \frac{n}{R_y}(t_0 + y_0) + \varphi_1$ is simply $\chi_{1,0,n}$ evaluated at $v =  \coeff{1}{\sqrt{2}}(t_0+y_0)$.

Transforming to Brinkman coordinates $x^{\mu''} = (x^-, x^+, z^1, z^2, z^3,z^4)$ using   (\ref{coordtrf1}), one finds
\begin{equation}
\begin{aligned}
B ~\to~ \frac{(Q_1 Q_5)^\frac{1}{4} \, b\, \theta\,   r^{n-1}}{ (r^2 +a^2)^{\frac{1}{2} (n+2)}} \, \, dx^- \wedge\, &\bigg[ \Big( r\, \Gamma \, \cos  \big(\chi^{(0)}_{1,0,n} - \eta(r)\big)+ a\, n \, \sin  \big(\chi^{(0)}_{1,0,n} - \eta(r)\big) \Big)  \, dz^3   \\
& ~+~    \Big( r\, \Gamma \, \sin \big(\chi^{(0)}_{1,0,n} - \eta(r)\big)  -  a\, n \, \cos  \big(\chi^{(0)}_{1,0,n} - \eta(r)\big) \Big) \,    dz^4\bigg]\,,
\end{aligned}
\label{Blim2}
\end{equation}
where we have used (\ref{SSreg1}) to write $(Q_1 Q_5)^\frac{1}{4} =  \sqrt{a\, A\, R_y}$.

In looking at (\ref{Blim2}), one should recall that $\lambda = 2  x^-$ and that $\lambda$ is related to $r$ via (\ref{rlambdareln}) and hence 
\begin{equation}
\frac{d r}{d x^-}  ~=~ -  A \,\Gamma \,.
  \label{rxminus}
\end{equation}
In addition, one should note that $\chi^{(0)}_{1,0,n}$ contains a $\varphi_1$ term and so  (\ref{Blim2}) contains terms linear in $z^1$ and $z^2$ that arise from 
$(Q_1 Q_5)^\frac{1}{4} \,  \theta \cos \chi^{(0)}_{1,0,n}$ and $(Q_1 Q_5)^\frac{1}{4} \, \theta \sin \chi^{(0)}_{1,0,n}$ via (\ref{trivtrf}).  It follows that the $3$-form, $H$, takes the form 
\begin{equation}
H ~=~d x^-  \, \wedge \, \sum_{a,b=1}^2 \, h_{a,b} (x^-)\, \, dz^a \wedge dz^{b+2}   \,,
  \label{Hform}
\end{equation}
for some functions, $ h_{a,b}$.

\end{appendix}

\newpage
\begin{adjustwidth}{-1mm}{-1mm} 
\bibliographystyle{utphys}      
\bibliography{microstates}       

\providecommand{\href}[2]{#2}\begingroup\raggedright\begin{thebibliography}{10}

\bibitem{Maldacena:1996ky}
J.~M. Maldacena, ``{Black holes in string theory},''
\href{http://arxiv.org/abs/hep-th/9607235}{{\ttfamily arXiv:hep-th/9607235}}.

\bibitem{Horowitz:1996rn}
G.~T. Horowitz, ``{Quantum states of black holes},'' in {\em {Symposium on
  Black Holes and Relativistic Stars (dedicated to memory of S.
  Chandrasekhar)}}, pp.~241--266.
\newblock 12, 1996.
\newblock \href{http://arxiv.org/abs/gr-qc/9704072}{{\ttfamily
  arXiv:gr-qc/9704072}}.

\bibitem{Martinec:1999bf}
E.~J. Martinec, ``{Black holes and the phases of brane thermodynamics},'' in
  {\em {NATO Advanced Study Institute: TMR Summer School on Progress in String
  Theory and M-Theory}}, pp.~117--145.
\newblock 5, 1999.
\newblock \href{http://arxiv.org/abs/hep-th/9909049}{{\ttfamily
  arXiv:hep-th/9909049}}.

\bibitem{Peet:2000hn}
A.~W. Peet, ``{TASI lectures on black holes in string theory},''
\href{http://arxiv.org/abs/hep-th/0008241}{{\ttfamily arXiv:hep-th/0008241}}.

\bibitem{Mathur:2005zp}
S.~D. Mathur, ``{The fuzzball proposal for black holes: An elementary
  review},'' \href{http://dx.doi.org/10.1002/prop.200410203}{{\em Fortsch.
  Phys.} {\bfseries 53} (2005) 793--827},
\href{http://arxiv.org/abs/hep-th/0502050}{{\ttfamily arXiv:hep-th/0502050}}.

\bibitem{Bena:2007kg}
I.~Bena and N.~P. Warner, ``{Black holes, black rings and their microstates},''
  \href{http://dx.doi.org/10.1007/978-3-540-79523-0}{{\em Lect. Notes Phys.}
  {\bfseries 755} (2008) 1--92},
\href{http://arxiv.org/abs/hep-th/0701216}{{\ttfamily arXiv:hep-th/0701216}}.

\bibitem{Bena:2013dka}
I.~Bena and N.~P. Warner, ``{Resolving the Structure of Black Holes:
  Philosophizing with a Hammer},''
\href{http://arxiv.org/abs/1311.4538}{{\ttfamily arXiv:1311.4538 [hep-th]}}.

\bibitem{Maldacena:1996ya}
J.~M. Maldacena, ``{Statistical entropy of near extremal five-branes},''
  \href{http://dx.doi.org/10.1016/0550-3213(96)00368-9}{{\em Nucl. Phys.}
  {\bfseries B477} (1996) 168--174},
\href{http://arxiv.org/abs/hep-th/9605016}{{\ttfamily arXiv:hep-th/9605016
  [hep-th]}}.

\bibitem{Horowitz:1996ay}
G.~T. Horowitz, J.~M. Maldacena, and A.~Strominger, ``{Nonextremal black hole
  microstates and U duality},''
  \href{http://dx.doi.org/10.1016/0370-2693(96)00738-1}{{\em Phys. Lett. B}
  {\bfseries 383} (1996) 151--159},
  \href{http://arxiv.org/abs/hep-th/9603109}{{\ttfamily arXiv:hep-th/9603109}}.

\bibitem{Danielsson:2001xe}
U.~H. Danielsson, A.~Guijosa, and M.~Kruczenski, ``{Brane anti-brane systems at
  finite temperature and the entropy of black branes},''
  \href{http://dx.doi.org/10.1088/1126-6708/2001/09/011}{{\em JHEP} {\bfseries
  09} (2001) 011}, \href{http://arxiv.org/abs/hep-th/0106201}{{\ttfamily
  arXiv:hep-th/0106201}}.

\bibitem{Banks:1997hz}
T.~Banks, W.~Fischler, I.~R. Klebanov, and L.~Susskind, ``{Schwarzschild black
  holes from matrix theory},''
  \href{http://dx.doi.org/10.1103/PhysRevLett.80.226}{{\em Phys. Rev. Lett.}
  {\bfseries 80} (1998) 226--229},
  \href{http://arxiv.org/abs/hep-th/9709091}{{\ttfamily arXiv:hep-th/9709091}}.

\bibitem{Banks:1997tn}
T.~Banks, W.~Fischler, I.~R. Klebanov, and L.~Susskind, ``{Schwarzschild black
  holes in matrix theory. 2.},''
  \href{http://dx.doi.org/10.1088/1126-6708/1998/01/008}{{\em JHEP} {\bfseries
  01} (1998) 008}, \href{http://arxiv.org/abs/hep-th/9711005}{{\ttfamily
  arXiv:hep-th/9711005}}.

\bibitem{Horowitz:1997fr}
G.~T. Horowitz and E.~J. Martinec, ``{Comments on black holes in matrix
  theory},'' \href{http://dx.doi.org/10.1103/PhysRevD.57.4935}{{\em Phys. Rev.
  D} {\bfseries 57} (1998) 4935--4941},
  \href{http://arxiv.org/abs/hep-th/9710217}{{\ttfamily arXiv:hep-th/9710217}}.

\bibitem{Horowitz:1996nw}
G.~T. Horowitz and J.~Polchinski, ``{A correspondence principle for black holes
  and strings},'' \href{http://dx.doi.org/10.1103/PhysRevD.55.6189}{{\em Phys.
  Rev.} {\bfseries D55} (1997) 6189--6197},
\href{http://arxiv.org/abs/hep-th/9612146}{{\ttfamily arXiv:hep-th/9612146}}.

\bibitem{Damour:1999aw}
T.~Damour and G.~Veneziano, ``{Selfgravitating fundamental strings and black
  holes},'' \href{http://dx.doi.org/10.1016/S0550-3213(99)00596-9}{{\em Nucl.
  Phys.} {\bfseries B568} (2000) 93--119},
\href{http://arxiv.org/abs/hep-th/9907030}{{\ttfamily arXiv:hep-th/9907030
  [hep-th]}}.

\bibitem{Bena:2004wt}
I.~Bena and P.~Kraus, ``{Three Charge Supertubes and Black Hole Hair},''
  \href{http://dx.doi.org/10.1103/PhysRevD.70.046003}{{\em Phys. Rev.}
  {\bfseries D70} (2004) 046003},
\href{http://arxiv.org/abs/hep-th/0402144}{{\ttfamily arXiv:hep-th/0402144}}.

\bibitem{Bena:2004wv}
I.~Bena, ``{Splitting hairs of the three charge black hole},''
  \href{http://dx.doi.org/10.1103/PhysRevD.70.105018}{{\em Phys. Rev.}
  {\bfseries D70} (2004) 105018},
\href{http://arxiv.org/abs/hep-th/0404073}{{\ttfamily arXiv:hep-th/0404073}}.

\bibitem{Gibbons:2013tqa}
G.~Gibbons and N.~Warner, ``{Global structure of five-dimensional fuzzballs},''
  \href{http://dx.doi.org/10.1088/0264-9381/31/2/025016}{{\em
  Class.Quant.Grav.} {\bfseries 31} (2014) 025016},
\href{http://arxiv.org/abs/1305.0957}{{\ttfamily arXiv:1305.0957 [hep-th]}}.

\bibitem{Tyukov:2017uig}
A.~Tyukov, R.~Walker, and N.~P. Warner, ``{Tidal Stresses and Energy Gaps in
  Microstate Geometries},''
  \href{http://dx.doi.org/10.1007/JHEP02(2018)122}{{\em JHEP} {\bfseries 02}
  (2018) 122},
\href{http://arxiv.org/abs/1710.09006}{{\ttfamily arXiv:1710.09006 [hep-th]}}.

\bibitem{Raju:2018xue}
S.~Raju and P.~Shrivastava, ``{A Critique of the Fuzzball Program},'' {\em
  Phys. Rev.} {\bfseries D99} no.~6, (2019) 066009,
\href{http://arxiv.org/abs/1804.10616}{{\ttfamily arXiv:1804.10616 [hep-th]}}.

\bibitem{Bianchi:2018kzy}
M.~Bianchi, D.~Consoli, A.~Grillo, and J.~F. Morales, ``{The dark side of
  fuzzball geometries},'' \href{http://dx.doi.org/10.1007/JHEP05(2019)126}{{\em
  JHEP} {\bfseries 05} (2019) 126},
  \href{http://arxiv.org/abs/1811.02397}{{\ttfamily arXiv:1811.02397
  [hep-th]}}.

\bibitem{Bena:2018mpb}
I.~Bena, E.~J. Martinec, R.~Walker, and N.~P. Warner, ``{Early Scrambling and
  Capped BTZ Geometries},''
\href{http://arxiv.org/abs/1812.05110}{{\ttfamily arXiv:1812.05110 [hep-th]}}.

\bibitem{Bena:2019azk}
I.~Bena, P.~Heidmann, R.~Monten, and N.~P. Warner, ``{Thermal Decay without
  Information Loss in Horizonless Microstate Geometries},''
\href{http://arxiv.org/abs/1905.05194}{{\ttfamily arXiv:1905.05194 [hep-th]}}.

\bibitem{Bena:2020iyw}
I.~Bena, A.~Houppe, and N.~P. Warner, ``{Delaying the Inevitable: Tidal
  Disruption in Microstate Geometries},''
  \href{http://arxiv.org/abs/2006.13939}{{\ttfamily arXiv:2006.13939
  [hep-th]}}.

\bibitem{Eperon:2016cdd}
F.~C. Eperon, H.~S. Reall, and J.~E. Santos, ``{Instability of supersymmetric
  microstate geometries},''
  \href{http://dx.doi.org/10.1007/JHEP10(2016)031}{{\em JHEP} {\bfseries 10}
  (2016) 031},
\href{http://arxiv.org/abs/1607.06828}{{\ttfamily arXiv:1607.06828 [hep-th]}}.

\bibitem{Marolf:2016nwu}
D.~Marolf, B.~Michel, and A.~Puhm, ``{A rough end for smooth microstate
  geometries},'' \href{http://dx.doi.org/10.1007/JHEP05(2017)021}{{\em JHEP}
  {\bfseries 05} (2017) 021},
\href{http://arxiv.org/abs/1612.05235}{{\ttfamily arXiv:1612.05235 [hep-th]}}.

\bibitem{Horowitz:1990sr}
G.~T. Horowitz and A.~R. Steif, ``{Strings in Strong Gravitational Fields},''
\href{http://dx.doi.org/10.1103/PhysRevD.42.1950}{{\em Phys. Rev.} {\bfseries
  D42} (1990) 1950--1959}.

\bibitem{Bena:2017upb}
I.~Bena, D.~Turton, R.~Walker, and N.~P. Warner, ``{Integrability and
  Black-Hole Microstate Geometries},''
  \href{http://dx.doi.org/10.1007/JHEP11(2017)021}{{\em JHEP} {\bfseries 11}
  (2017) 021},
\href{http://arxiv.org/abs/1709.01107}{{\ttfamily arXiv:1709.01107 [hep-th]}}.

\bibitem{Heidmann:2019zws}
P.~Heidmann and N.~P. Warner, ``{Superstratum Symbiosis},''
\href{http://arxiv.org/abs/1903.07631}{{\ttfamily arXiv:1903.07631 [hep-th]}}.

\bibitem{Heidmann:2019xrd}
P.~Heidmann, D.~R. Mayerson, R.~Walker, and N.~P. Warner, ``{Holomorphic Waves
  of Black Hole Microstructure},''
  \href{http://dx.doi.org/10.1007/JHEP02(2020)192}{{\em JHEP} {\bfseries 02}
  (2020) 192}, \href{http://arxiv.org/abs/1910.10714}{{\ttfamily
  arXiv:1910.10714 [hep-th]}}.

\bibitem{Mayerson:2020tcl}
D.~R. Mayerson, R.~A. Walker, and N.~P. Warner, ``{Microstate Geometries from
  Gauged Supergravity in Three Dimensions},''
  \href{http://arxiv.org/abs/2004.13031}{{\ttfamily arXiv:2004.13031
  [hep-th]}}.

\bibitem{Bena:2017xbt}
I.~Bena, S.~Giusto, E.~J. Martinec, R.~Russo, M.~Shigemori, D.~Turton, and
  N.~P. Warner, ``{Asymptotically-flat supergravity solutions deep inside the
  black-hole regime},'' \href{http://dx.doi.org/10.1007/JHEP02(2018)014}{{\em
  JHEP} {\bfseries 02} (2018) 014},
\href{http://arxiv.org/abs/1711.10474}{{\ttfamily arXiv:1711.10474 [hep-th]}}.

\bibitem{Blau:2002mw}
M.~Blau, J.~M. Figueroa-O'Farrill, and G.~Papadopoulos, ``{Penrose limits,
  supergravity and brane dynamics},''
  \href{http://dx.doi.org/10.1088/0264-9381/19/18/310}{{\em Class. Quant.
  Grav.} {\bfseries 19} (2002) 4753},
\href{http://arxiv.org/abs/hep-th/0202111}{{\ttfamily arXiv:hep-th/0202111
  [hep-th]}}.

\bibitem{Giusto:2013rxa}
S.~Giusto, L.~Martucci, M.~Petrini, and R.~Russo, ``{6D microstate geometries
  from 10D structures},''
  \href{http://dx.doi.org/10.1016/j.nuclphysb.2013.08.018}{{\em Nucl.Phys.}
  {\bfseries B876} (2013) 509--555},
\href{http://arxiv.org/abs/1306.1745}{{\ttfamily arXiv:1306.1745 [hep-th]}}.

\bibitem{Bena:2015bea}
I.~Bena, S.~Giusto, R.~Russo, M.~Shigemori, and N.~P. Warner, ``{Habemus
  Superstratum! A constructive proof of the existence of superstrata},''
  \href{http://dx.doi.org/10.1007/JHEP05(2015)110}{{\em JHEP} {\bfseries 05}
  (2015) 110},
\href{http://arxiv.org/abs/1503.01463}{{\ttfamily arXiv:1503.01463 [hep-th]}}.

\bibitem{Russo:2002qj}
J.~G. Russo and A.~A. Tseytlin, ``{A Class of exact pp wave string models with
  interacting light cone gauge actions},''
  \href{http://dx.doi.org/10.1088/1126-6708/2002/09/035}{{\em JHEP} {\bfseries
  09} (2002) 035}, \href{http://arxiv.org/abs/hep-th/0208114}{{\ttfamily
  arXiv:hep-th/0208114}}.

\bibitem{Mizoguchi:2002qy}
S.~Mizoguchi, T.~Mogami, and Y.~Satoh, ``{Penrose limits and Green-Schwarz
  strings},'' \href{http://dx.doi.org/10.1088/0264-9381/20/8/306}{{\em Class.
  Quant. Grav.} {\bfseries 20} (2003) 1489--1502},
  \href{http://arxiv.org/abs/hep-th/0209043}{{\ttfamily arXiv:hep-th/0209043}}.

\bibitem{Wands:2007bd}
D.~Wands, ``{Multiple field inflation},''
  \href{http://dx.doi.org/10.1007/978-3-540-74353-8\_8}{{\em Lect. Notes Phys.}
  {\bfseries 738} (2008) 275--304},
  \href{http://arxiv.org/abs/astro-ph/0702187}{{\ttfamily
  arXiv:astro-ph/0702187}}.

\bibitem{Langlois:2010xc}
D.~Langlois, ``{Lectures on inflation and cosmological perturbations},''
  \href{http://dx.doi.org/10.1007/978-3-642-10598-2\_1}{{\em Lect. Notes in
  Phys.} {\bfseries 800} (2010) 1--57},
  \href{http://arxiv.org/abs/1001.5259}{{\ttfamily arXiv:1001.5259
  [astro-ph.CO]}}.

\bibitem{DaCunha:2003fm}
B.~Carneiro~da Cunha and E.~J. Martinec, ``{Closed string tachyon condensation
  and world sheet inflation},''
  \href{http://dx.doi.org/10.1103/PhysRevD.68.063502}{{\em Phys. Rev. D}
  {\bfseries 68} (2003) 063502},
  \href{http://arxiv.org/abs/hep-th/0303087}{{\ttfamily arXiv:hep-th/0303087}}.

\bibitem{Martinec:2014uva}
E.~J. Martinec and W.~E. Moore, ``{Modeling Quantum Gravity Effects in
  Inflation},'' \href{http://dx.doi.org/10.1007/JHEP07(2014)053}{{\em JHEP}
  {\bfseries 07} (2014) 053}, \href{http://arxiv.org/abs/1401.7681}{{\ttfamily
  arXiv:1401.7681 [hep-th]}}.

\bibitem{Audretsch:1979uv}
J.~Audretsch, ``{COSMOLOGICAL PARTICLE CREATION AS ABOVE - BARRIER REFLECTION:
  APPROXIMATION METHOD AND APPLICATIONS},''
  \href{http://dx.doi.org/10.1088/0305-4470/12/8/013}{{\em J. Phys. A}
  {\bfseries 12} (1979) 1189--1203}.

\bibitem{Lawrence:1995ct}
A.~E. Lawrence and E.~J. Martinec, ``{String field theory in curved space-time
  and the resolution of space - like singularities},''
  \href{http://dx.doi.org/10.1088/0264-9381/13/1/007}{{\em Class. Quant. Grav.}
  {\bfseries 13} (1996) 63--96},
  \href{http://arxiv.org/abs/hep-th/9509149}{{\ttfamily arXiv:hep-th/9509149}}.

\bibitem{Silverstein:2014yza}
E.~Silverstein, ``{Backdraft: String Creation in an Old Schwarzschild Black
  Hole},'' \href{http://arxiv.org/abs/1402.1486}{{\ttfamily arXiv:1402.1486
  [hep-th]}}.

\bibitem{Eckart:1930zza}
C.~Eckart, ``{The Penetration of a Potential Barrier by Electrons},''
  \href{http://dx.doi.org/10.1103/PhysRev.35.1303}{{\em Phys. Rev.} {\bfseries
  35} (1930) 1303--1309}.

\bibitem{Larsen:1999uk}
F.~Larsen and E.~J. Martinec, ``{U(1) charges and moduli in the D1-D5
  system},'' {\em JHEP} {\bfseries 06} (1999) 019,
\href{http://arxiv.org/abs/hep-th/9905064}{{\ttfamily arXiv:hep-th/9905064}}.

\bibitem{Albrecht:1992kf}
A.~Albrecht, P.~Ferreira, M.~Joyce, and T.~Prokopec, ``{Inflation and squeezed
  quantum states},'' \href{http://dx.doi.org/10.1103/PhysRevD.50.4807}{{\em
  Phys. Rev. D} {\bfseries 50} (1994) 4807--4820},
  \href{http://arxiv.org/abs/astro-ph/9303001}{{\ttfamily
  arXiv:astro-ph/9303001}}.

\bibitem{Bena:2018bbd}
I.~Bena, P.~Heidmann, and D.~Turton, ``{AdS$_{2}$ holography: mind the cap},''
  \href{http://dx.doi.org/10.1007/JHEP12(2018)028}{{\em JHEP} {\bfseries 12}
  (2018) 028},
\href{http://arxiv.org/abs/1806.02834}{{\ttfamily arXiv:1806.02834 [hep-th]}}.

\bibitem{Martinec:2020gkv}
E.~J. Martinec, S.~Massai, and D.~Turton, ``{Stringy Structure at the BPS
  Bound},'' \href{http://arxiv.org/abs/2005.12344}{{\ttfamily arXiv:2005.12344
  [hep-th]}}.

\bibitem{Marolf:2011dj}
D.~Marolf and A.~Ori, ``{Outgoing gravitational shock-wave at the inner
  horizon: The late-time limit of black hole interiors},''
  \href{http://dx.doi.org/10.1103/PhysRevD.86.124026}{{\em Phys. Rev.}
  {\bfseries D86} (2012) 124026},
\href{http://arxiv.org/abs/1109.5139}{{\ttfamily arXiv:1109.5139 [gr-qc]}}.

\bibitem{Eilon:2016osg}
E.~Eilon and A.~Ori, ``{Numerical study of the gravitational shock wave inside
  a spherical charged black hole},''
  \href{http://dx.doi.org/10.1103/PhysRevD.94.104060}{{\em Phys. Rev. D}
  {\bfseries 94} no.~10, (2016) 104060},
  \href{http://arxiv.org/abs/1610.04355}{{\ttfamily arXiv:1610.04355 [gr-qc]}}.

\bibitem{Chesler:2018hgn}
P.~M. Chesler, E.~Curiel, and R.~Narayan, ``{Numerical evolution of shocks in
  the interior of Kerr black holes},''
  \href{http://dx.doi.org/10.1103/PhysRevD.99.084033}{{\em Phys. Rev. D}
  {\bfseries 99} no.~8, (2019) 084033},
  \href{http://arxiv.org/abs/1808.07502}{{\ttfamily arXiv:1808.07502 [gr-qc]}}.

\bibitem{Burko:2019fgt}
L.~M. Burko and G.~Khanna, ``{Marolf-Ori singularity inside fast spinning black
  holes},'' \href{http://dx.doi.org/10.1103/PhysRevD.99.081501}{{\em Phys. Rev.
  D} {\bfseries 99} no.~8, (2019) 081501},
  \href{http://arxiv.org/abs/1901.03413}{{\ttfamily arXiv:1901.03413 [gr-qc]}}.

\bibitem{Dias:2019ery}
O.~J. Dias, H.~S. Reall, and J.~E. Santos, ``{The BTZ black hole violates
  strong cosmic censorship},''
  \href{http://dx.doi.org/10.1007/JHEP12(2019)097}{{\em JHEP} {\bfseries 12}
  (2019) 097}, \href{http://arxiv.org/abs/1906.08265}{{\ttfamily
  arXiv:1906.08265 [hep-th]}}.

\bibitem{Pandya:2020ejc}
A.~Pandya and F.~Pretorius, ``{The rotating black hole interior: Insights from
  gravitational collapse in $AdS_3$ spacetime},''
  \href{http://dx.doi.org/10.1103/PhysRevD.101.104026}{{\em Phys. Rev. D}
  {\bfseries 101} no.~10, (2020) 104026},
  \href{http://arxiv.org/abs/2002.07130}{{\ttfamily arXiv:2002.07130 [gr-qc]}}.

\bibitem{Emparan:2020rnp}
R.~Emparan and M.~Tomasevi\'c, ``{Strong cosmic censorship in the BTZ black
  hole},'' \href{http://dx.doi.org/10.1007/JHEP06(2020)038}{{\em JHEP}
  {\bfseries 06} (2020) 038}, \href{http://arxiv.org/abs/2002.02083}{{\ttfamily
  arXiv:2002.02083 [hep-th]}}.

\bibitem{Callan:1996dv}
C.~G. Callan and J.~M. Maldacena, ``{D-brane Approach to Black Hole Quantum
  Mechanics},'' \href{http://dx.doi.org/10.1016/0550-3213(96)00225-8}{{\em
  Nucl. Phys.} {\bfseries B472} (1996) 591--610},
\href{http://arxiv.org/abs/hep-th/9602043}{{\ttfamily arXiv:hep-th/9602043}}.

\bibitem{Das:1996wn}
S.~R. Das and S.~D. Mathur, ``{Comparing decay rates for black holes and
  D-branes},'' \href{http://dx.doi.org/10.1016/0550-3213(96)00453-1}{{\em Nucl.
  Phys.} {\bfseries B478} (1996) 561--576},
\href{http://arxiv.org/abs/hep-th/9606185}{{\ttfamily arXiv:hep-th/9606185}}.

\bibitem{Damour:2001bk}
T.~Damour and A.~Vilenkin, ``{Gravitational wave bursts from cusps and kinks on
  cosmic strings},'' \href{http://dx.doi.org/10.1103/PhysRevD.64.064008}{{\em
  Phys. Rev. D} {\bfseries 64} (2001) 064008},
  \href{http://arxiv.org/abs/gr-qc/0104026}{{\ttfamily arXiv:gr-qc/0104026}}.

\bibitem{Martinec:2018nco}
E.~J. Martinec, S.~Massai, and D.~Turton, ``{String dynamics in NS5-F1-P
  geometries},'' \href{http://dx.doi.org/10.1007/JHEP09(2018)031}{{\em JHEP}
  {\bfseries 09} (2018) 031},
\href{http://arxiv.org/abs/1803.08505}{{\ttfamily arXiv:1803.08505 [hep-th]}}.

\bibitem{Shigemori:2019orj}
M.~Shigemori, ``{Counting Superstrata},''
  \href{http://dx.doi.org/10.1007/JHEP10(2019)017}{{\em JHEP} {\bfseries 10}
  (2019) 017},
\href{http://arxiv.org/abs/1907.03878}{{\ttfamily arXiv:1907.03878 [hep-th]}}.

\bibitem{Strominger:1996sh}
A.~Strominger and C.~Vafa, ``{Microscopic Origin of the Bekenstein-Hawking
  Entropy},'' \href{http://dx.doi.org/10.1016/0370-2693(96)00345-0}{{\em Phys.
  Lett.} {\bfseries B379} (1996) 99--104},
\href{http://arxiv.org/abs/hep-th/9601029}{{\ttfamily arXiv:hep-th/9601029}}.

\bibitem{Martinec:2019wzw}
E.~J. Martinec, S.~Massai, and D.~Turton, ``{Little Strings, Long Strings, and
  Fuzzballs},'' \href{http://dx.doi.org/10.1007/JHEP11(2019)019}{{\em JHEP}
  {\bfseries 11} (2019) 019},
\href{http://arxiv.org/abs/1906.11473}{{\ttfamily arXiv:1906.11473 [hep-th]}}.

\bibitem{Bena:2020yii}
I.~Bena, F.~Eperon, P.~Heidmann, and N.~P. Warner, ``{The Great Escape:
  Tunneling out of Microstate Geometries},''
  \href{http://arxiv.org/abs/2005.11323}{{\ttfamily arXiv:2005.11323
  [hep-th]}}.

\bibitem{Cardoso:2019rvt}
V.~Cardoso and P.~Pani, ``{Testing the nature of dark compact objects: a status
  report},'' \href{http://dx.doi.org/10.1007/s41114-019-0020-4}{{\em Living
  Rev. Rel.} {\bfseries 22} no.~1, (2019) 4},
  \href{http://arxiv.org/abs/1904.05363}{{\ttfamily arXiv:1904.05363 [gr-qc]}}.

\bibitem{Abedi:2020sgg}
J.~Abedi and N.~Afshordi, ``{Echoes from the Abyss: A Status Update},''
  \href{http://arxiv.org/abs/2001.00821}{{\ttfamily arXiv:2001.00821 [gr-qc]}}.

\bibitem{Dodelson:2020lal}
M.~Dodelson and H.~Ooguri, ``{Singularities of thermal correlators at strong
  coupling},'' \href{http://arxiv.org/abs/2010.09734}{{\ttfamily
  arXiv:2010.09734 [hep-th]}}.

\bibitem{Shigemori:2013lta}
M.~Shigemori, ``{Perturbative 3-charge microstate geometries in six
  dimensions},'' \href{http://dx.doi.org/10.1007/JHEP10(2013)169}{{\em JHEP}
  {\bfseries 1310} (2013) 169},
\href{http://arxiv.org/abs/1307.3115}{{\ttfamily arXiv:1307.3115}}.

\bibitem{Bena:2016ypk}
I.~Bena, S.~Giusto, E.~J. Martinec, R.~Russo, M.~Shigemori, D.~Turton, and
  N.~P. Warner, ``{Smooth horizonless geometries deep inside the black-hole
  regime},'' \href{http://dx.doi.org/10.1103/PhysRevLett.117.201601}{{\em Phys.
  Rev. Lett.} {\bfseries 117} no.~20, (2016) 201601},
\href{http://arxiv.org/abs/1607.03908}{{\ttfamily arXiv:1607.03908 [hep-th]}}.

\bibitem{Gutowski:2003rg}
J.~B. Gutowski, D.~Martelli, and H.~S. Reall, ``{All supersymmetric solutions
  of minimal supergravity in six dimensions},''
  \href{http://dx.doi.org/10.1088/0264-9381/20/23/008}{{\em Class. Quant.
  Grav.} {\bfseries 20} (2003) 5049--5078},
\href{http://arxiv.org/abs/hep-th/0306235}{{\ttfamily arXiv:hep-th/0306235}}.

\end{thebibliography}\endgroup

\end{adjustwidth}


\end{document}